\documentclass[preprintnumbers,article,amsmath,amssymb,floatfix,10pt,prd,onecolumn,superscriptaddress,nofootinbib]{revtex4}
\usepackage[colorlinks=true, pdfstartview=FitV, linkcolor=magenta, citecolor=red, urlcolor=blue]{hyperref}
\usepackage{epsfig,color}
\usepackage{xcolor}
\usepackage{epstopdf}
\usepackage{amssymb}
\usepackage{verbatim}
\usepackage{multirow}
\usepackage{latexsym}
\usepackage{epsfig}
\usepackage{epstopdf}
\usepackage{graphicx}
\usepackage{subcaption}
\usepackage{amssymb}
\usepackage{amsmath}
\usepackage{dcolumn}
\usepackage{float}
\usepackage{bm}
\usepackage{color}
\usepackage{comment}
\usepackage{orcidlink}
\begin{document}

\title{Thermodynamic topology and orbital dynamics of charged RN--AdS black hole with logarithmic corrections}



\author{Hafiza Bushra Mujahid}
\email{bushrashahzad272@gmail.com}
\affiliation{Department of Mathematics, COMSATS University Islamabad, Sahiwal Campus-57000, Islamabad, Pakistan}
\author{Muhammad Yasir \orcidlink{0000-0003-0197-1725}}
\email{yasirciitsahiwal@gmail.com; muhammadyasir@nbu.edu.cn (Corresponding author)}
\affiliation{Institute of Fundamental Physics and Quantum Technology, Department of Physics, School of Physical Science and Technology, Ningbo University, Ningbo, Zhejiang 315211, People's Republic of China}
\affiliation{Center for Theoretical Physics, Khazar University, 41 Mehseti Str., Baku, AZ1096, Azerbaijan}
\author{Shahid Qaisar}
\email{shahidqaisar90@ciitsahiwal.edu.pk}
\affiliation{Department of Mathematics, COMSATS University Islamabad, Sahiwal Campus-57000, Islamabad, Pakistan}
\author{Faisal Javed}
\email{faisaljaved.math@gmail.com}
\affiliation{Department of Physics, Zhejiang Normal University, Jinhua 321004, People's Republic of China}

\begin{abstract}

This study investigates the topological thermodynamic properties and orbital dynamics of the Reissner--Nordstr\"{o}m anti-de Sitter (RN--AdS) black hole with logarithmic corrections solution in Einstein--Maxwell gravity. By utilizing the Duan's $\Phi$-mapping theory, we investigate the topological classes within the generalized three-parameter Sharma--Mittal entropy formalism.
Employing different thermodynamic ensembles, including the canonical and mixed canonical ensembles, we explore the critical, divergence points and heat capacity, which help us determine the thermodynamic stability of the BH. In addition, the Hamiltonian formalism is employed to investigate the orbital and oscillatory motion of neutral test particles. By utilizing the effective potential approach, we examine the effective force, the stability of circular orbits, and the innermost stable circular orbit (ISCO). We further analyze the trajectories of test particles and derive the radial $\left(\omega_{r}\right)$, latitudinal $\left(\omega_{\theta}\right)$, and azimuthal $\left(\omega_{\phi}\right)$ oscillation frequencies, together with the periastron precession frequency.
This work provides key insights into the fundamental nature of Einstein--Maxwell gravity. Moreover, these results provide a comprehensive theoretical description of the orbital dynamics of the RN--AdS black hole solution and may be useful for future high-precision astrophysical observations.


\end{abstract}
\maketitle
\section{Introduction}

Black holes (BHs) are among the most interesting predictions of Einstein's General Relativity (GR), providing a systematic framework for investigating the fundamental nature of gravity, spacetime geometry, and the interplay between gravitational and quantum phenomena. Since the development of BH physics, considerable attention has been devoted to understanding their thermodynamic properties, particularly the relationships among mass, temperature, entropy, and horizon geometry. The formulation of the four laws of BH mechanics and the discovery of Hawking radiation established a profound connection between thermodynamics and the dynamics of gravity \cite{00,000,0,01,1,3,4,5}.
Within the conventional framework, the Bekenstein--Hawking area law states that the entropy of a BH is proportional to the area of its event horizon. Another important feature is that BHs possess a characteristic temperature, known as the Hawking temperature, which is inversely proportional to the mass of a non-rotating BH. These remarkable developments established BH thermodynamics as a fundamental area of modern theoretical physics.

Alternative entropy formalisms have also been proposed to investigate possible non-extensive effects associated with gravitational systems. In particular, non-extensive thermodynamics suggests that the definition and behavior of entropy may depend on the physical context and can differ across theoretical frameworks, thereby motivating a reconsideration of its conventional role \cite{6,7,12}. This has led to the exploration of alternative entropy models, including the R\'{e}nyi \cite{8,9,10} and Tsallis \cite{11} formulations, where non-additive entropy measures provide valuable insights into the thermodynamic behavior of non-extensive systems.

Further theoretical refinements in generalized entropy formalism have been introduced in recent years. Motivated by quantum nature of spacetime, Barrow entropy modifies the area law by introducing fractal corrections to the horizon geometry \cite{13,14,15,16}. Sharma-Mittal entropy bridges the forms of R\'{e}nyi entropy and Tsallis entropy establishing a flexible framework that reduces to the Bekenstein-Hawking result in specific limits, meanwhile, Kaniadakis entropy based on relativistic statistical mechanics, provides another avenue for capturing non-extensive effects \cite{17,18,19,20,21,22,23,24,25,26}. As pointed out in reference \cite{27, 28, TC1, TC2, TC3, TC4, TC5, TC6, TC7, TC8}, entropy corrections play a crucial role in determining the stability and phase transitions of BH thermodynamics and significantly influence physical quantities such as Gibbs free energy, Helmholtz free energy, and heat capacity. Within the framework of Sharma-Mittal entropy, which  gives a generalized framework that relates various non-extensive entropy measures by extra parameters, this paper focuses on the thermodynamic behavior of the RN--AdS black hole in the context of nonlinear electrodynamics (NED) , which is a charged and non-rotating solution of the Einstein–Maxwell equations  \cite{29,30,31,32,36}. 
Furthermore, the topological characteristics of the considered BH are investigated in different thermodynamic ensembles, including the canonical ensemble, in which the BH charge is fixed, and the mixed canonical ensemble, where additional fluctuations are taken into account, as discussed in Refs. \cite{33,34,35,37}. The topological classes of the system are analyzed within both the canonical and mixed canonical ensembles, allowing us to identify precise differences in thermodynamic stability \cite{Ensml1,Ensml2,Ensml3,Ensml4,Ensml5}. This formalism is based on Duan's \(\Phi\)-mapping theory, which relates the zeros of a suitably defined vector field to thermodynamic critical points and classifies their topological properties \cite{TC2,TC3,TC6,TC7,TC8,TC9,TC10,TC11,TC12,TC13,TC14,TC15}. Within this framework, the thermodynamic defects and corresponding winding numbers are investigated for different BH configurations. This framework offers great interest into the phase structure, stability boundaries, and critical behavior of charged BHs while providing a broader perspective on their micro-states and underlying statistical mechanics. Moreover, the specific heat capacity is effectively investigate the thermal stability of the BH, which clarify the understanding of the thermodynamic phase structure and critical phenomena of BHs by their sign convention \cite{40,41,42}. In particular, the positive heat capacity signifies that the BH is thermally stable, while the negative heat capacity signifies that the unstable thermodynamic configuration of the BH.

Particle dynamics is also an important aspect of the BH study, where the investigation of the particle motion around compact astrophysical objects provide valuable understanding about  hamiltonian formalism of charged BHs, orbital motion, epicyclic oscillation, and formation of ISCOs as discussed in Refs. \cite{OME,43,44,45}. The thermodynamic framework provide global spacetime properties, while the dynamics of the test particle provides a local geometric characteristics and stability of the gravitational backgrounds. In particular, the effective potential gives the conditions for circular motion and orbital stability by analyzing the first and second derivatives, respectively \cite{38,39,46}. This formal framework was initially established within the context of Schwarzschild and Kerr spacetimes and subsequently extended to a broad class of BH solutions involving charge, higher-order curvature corrections and NED. Previous investigations have demonstrated that electric charge, higher-curvature effects, and nonlinear electromagnetic corrections significantly influence the effective potential, ISCO radius, orbital stability, and epicyclic frequencies, highlighting the sensitivity of particle dynamics to the underlying spacetime parameters, including charge, spin, and coupling constants \cite{47,48,49,50}. In addition, small perturbations about stable circular orbits give information about the radial and the latitudinal epicyclic frequencies, which are important for gaining insight into orbital oscillations and periastron precession. Derivation of the equations of motion and examination of the orbital properties can be done systematically using the Hamiltonian formalism. Furthermore, the centre-of-mass energy of colliding particles is another convenient property to consider particle interactions in strong gravitational fields.

The main objective of this paper is to investigate how Sharma-Mittal entropy alters both the thermodynamic and dynamical behavior of a logarithmically corrected charged RN--AdS black hole, particularly regarding their stability conditions, critical phenomena, and phase transition structure. Since the standard Bekenstein-Hawking entropy may not fully capture possible non-extensive effects that arising in gravitational systems, Sharma-Mittal entropy offers an extended framework for exploring thermodynamic properties that go beyond the standard area law. We derive the generalized thermodynamic relations by including Sharma-Mittal entropy into the BH framework, ensuring that all relevant quantities including temperature, pressure, generalized free energy, heat capacity, and other thermodynamic potentials explicitly reflect the effects of the entropy parameters \cite{53,54}. We examine the stability of BH phases through critical point conditions and phase transition behavior. In addition to thermodynamic analysis, this paper investigates particle dynamics, aiming to understand how generalized entropy corrections relate to observable dynamical properties. Using the Hamiltonian formalism, this study analyzes the motion of test particles by examining effective potential profiles, circular orbits, conditions for orbital stability and energy angular momentum relationships. Furthermore, the study explores the harmonic oscillations as perturbation of circular orbits, centre of mass energy \cite{57}. Through graphical and numerical analyses, the study demonstrate how variations in the Sharma-Mittal entropy parameters focuses on identifying underlying thermodynamic topological structures and exploring how generalized entropy corrections effect the emergence of stable and unstable BH phases.

This study investigates BH topological charges within
the framework of three-parameter generalized entropy. We first provide an overview of a novel class of BHs with Sharma-Mittal entropy in Section \ref{A1} and examine the BH thermodynamic topology in the canonical ensemble in Section \ref{B}. Then a BH solution is presented as topological thermodynamic defects for both canonical in Section \ref{C} and mixed canonical ensembles in Sections \ref{D}, \ref{E}. Finally, we analyze the BH heat capacity in Section \ref{G}. In Section \ref{K}, we discuss about particle dynamics and discuss its effective potential in Section \ref{H}. In Sections \ref{Q}, \ref{I}, \ref{HM}, \ref{N}, we discuss about ISCO, effective force, Harmonic oscillations and center of mass energy respectively. Finally, we provide concluding remarks in Section \ref{J} summarizing the key findings of this study.

\section{A brief description of BH and three parameter generalized Entropy}\label{A1}

For NED theories minimally coupled to gravity, the action can be written as \cite{36}
\begin{equation}
\mathbb{S}=\frac{1}{16\pi} \int \sqrt{-g} \big[R+K(\mathcal{F})\big]
d^4x,
\end{equation}
where $\mathcal{F} = F_{\mu\nu}F^{\mu\nu},$ and $ F_{\mu\nu} = \nabla_\mu A_\nu-\nabla_\nu A_\mu .$ Here, $A_\mu$ represents the Maxwell field, and $R$ denotes the Ricci scalar. The function $K(\mathcal{F})$ depends on the electromagnetic invariant $\mathcal{F}$. Varying the action with respect to the metric yields the Einstein field equations:
\begin{equation}
G_{\mu\nu} = -2K_{,\mathcal{F}} F_{\mu\lambda}F_{\nu}^{\;\;\lambda}
+ \tfrac{1}{2} g_{\mu\nu} K , \qquad K_{,\mathcal{F}} \equiv
\frac{dK}{d\mathcal{F}} ,
\end{equation}
and the variation of the action with respect to the gauge field $A_\mu$, one derives the generalized Maxwell equations: $\nabla_\mu \left( K_{,\mathcal{F}} F^{\mu\nu} \right) = 0 .$ In a static and spherically symmetric spacetime context, the line element can always be parameterized as
\begin{equation}
ds^2 = -U(r)\,dt^2 + \frac{dr^2}{U(r)} + r^2 \, d\Omega_2^2 ,
\end{equation}
with $d\Omega_2^2 = d\theta^2 + \sin^2\theta \, d\phi^2.$ Here, the metric function $U(r)$ defined in \cite{36}, which is described as
\begin{equation}\label{1}
U(r) = 1 - \frac{2M}{r} + \frac{Q^{2}}{r^{2}} - \frac{1}{3} \Lambda
r^{2} - \frac{2M'}{r},
\end{equation}
with
\begin{equation}\nonumber\label{2}
M' \equiv -\frac{1}{3} \, \frac{Q^{2} \, \ln\!\left( 16 \zeta Q
\right)}{\sqrt {\zeta Q}}.
\end{equation}

\begin{figure}\center
\epsfig{file=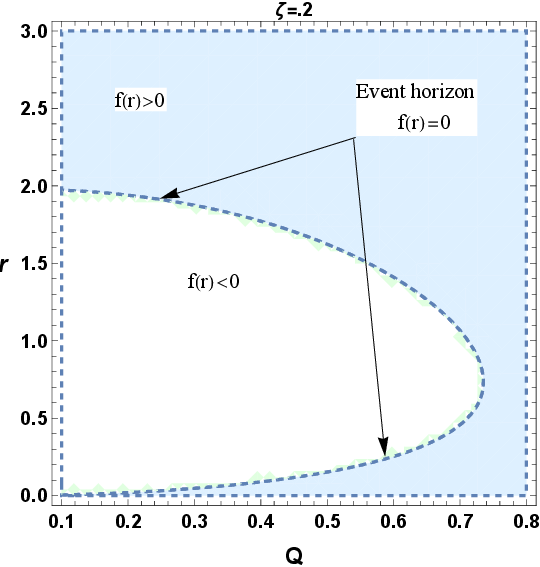,width=.3\linewidth,height=2.0in}\quad\epsfig{file=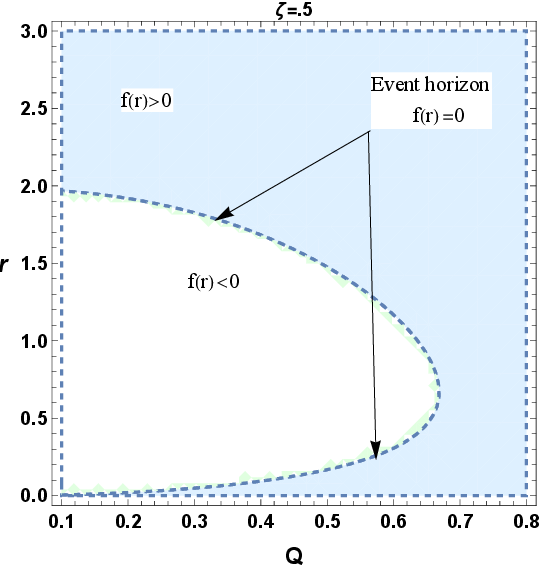,width=.3\linewidth,height=2.0in}\quad\epsfig{file=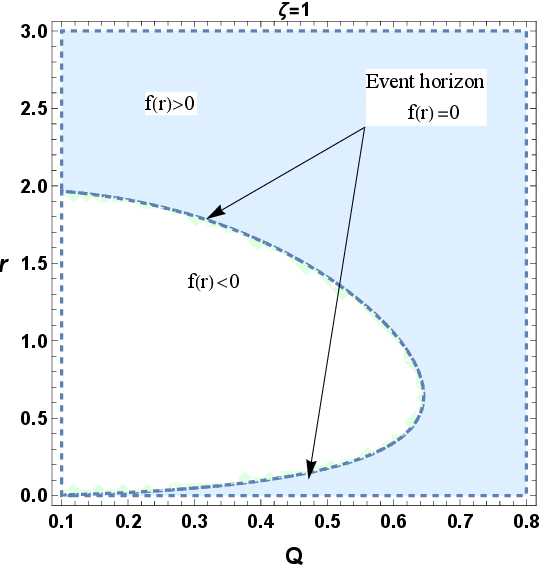,width=.3\linewidth,height=2.0in}
\caption{\label{et1} Behavior of lapse function for different values $\zeta$ along radial coordinate and charge with $M=1$. The boundary between these regions expresses the position of the horizon radius where the lapse function vanishes.}
\epsfig{file=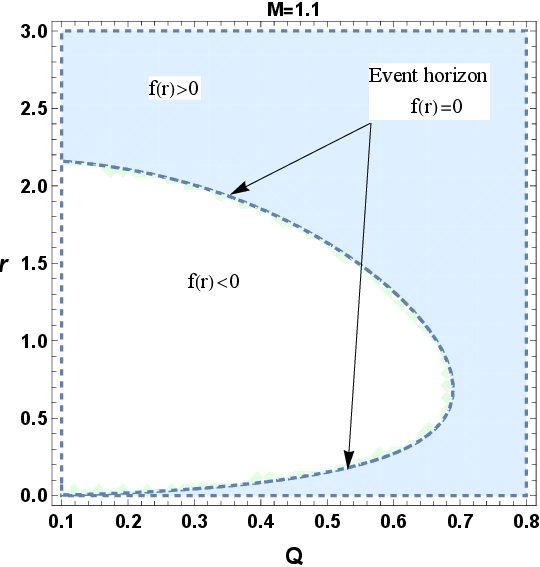,width=.3\linewidth,height=2.0in}\quad\epsfig{file=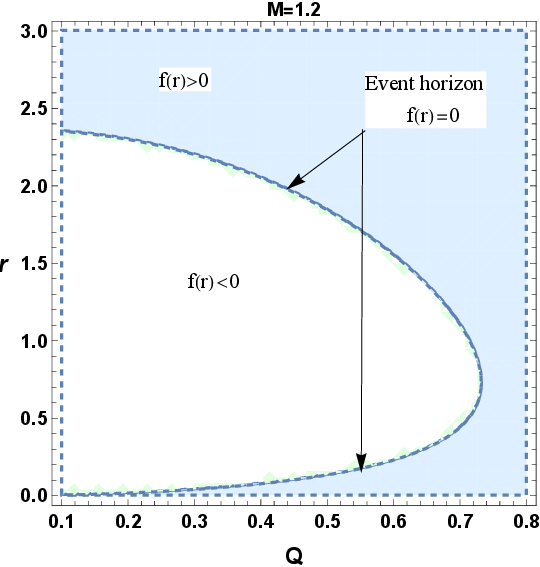,width=.3\linewidth,height=2.0in}\quad\epsfig{file=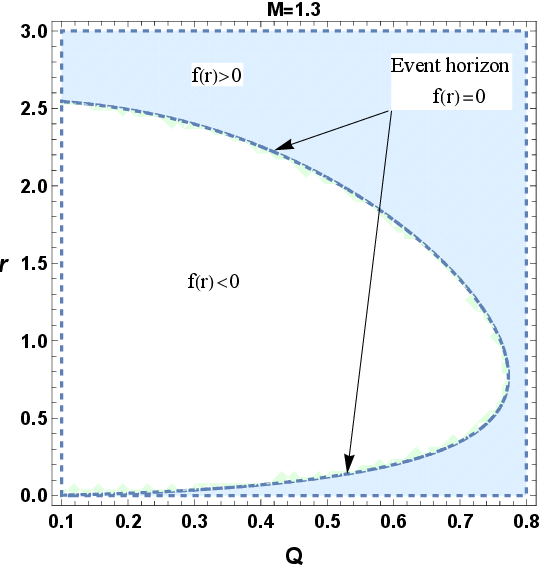,width=.3\linewidth,height=2.0in}
\caption{\label{et2} Behavior of lapse function for different values $M$ along radial coordinate and charge with $\zeta=1$. }
\end{figure}
A small coupling constant $\zeta$ effectively contributes as an
effective mass. For an extremal BH, we have $Q = M + M',$ indicating that the electric charge $Q$ can significantly exceed the BH mass $M$ in the absence of the coupling parameter $\zeta$. The impact of BH parameters on the position of horizon radius is presented in Figs. (\ref{et1}) and (\ref{et2}). The plots illustrate the behavior of the lapse function $(U(r))$, whose zeros determine the location of the event horizon and therefore characterize the causal structure of the spacetime. The boundary separating the regions $(U(r)>0)$ and $(U(r)<0)$ corresponds to $(U(r)=0)$, indicating the formation of a BH horizon. Increasing the mass parameter $(M)$ enlarges the domain of positive lapse function, leading to an outward displacement of the event horizon. For the pressure convention, the thermodynamic pressure $P$ of a BH is related to the cosmological constant $\Lambda$ and the AdS length scale parameter $l$ from Ref. \cite{Ref2, Ref0}, and the characteristic length scale parameter $l$ as:
\begin{equation}\label{3}
P = \frac{3}{8 \pi l^{2}} = -\frac{\Lambda}{8 \pi G} .
\end{equation}
where \(G\) denotes the gravitational constant $(G=1)$. The BH entropy at the horizon becomes $S_0= {\pi r_{h}^2}$. The BH mass $M$ takes the form
\begin{equation}
\label{M}
M=\frac{1}{6} \left(\frac{S (8 P S+3)+3 \pi  Q^2}{\sqrt{\pi } \sqrt{S}}+Q^2 \log (256 \zeta  Q)\right).
\end{equation}
The first law of thermodynamics is defined by
\begin{equation}
dM=VdP+TdS+Xd\zeta +\chi dQ,
\end{equation}
with $V$ (thermodynamic BH volume), $\chi$ (electrostatic potential), the conjugate factor $X$ and the temperature $T$ are expressed as
\begin{equation}\label{SV}
V=\frac{4 S^{3/2}}{3 \sqrt{\pi }}, \quad \chi=\frac{1}{6} Q \left(\frac{2}{\sqrt{\zeta  Q}}+\frac{3 \log (16 \zeta  Q)}{\sqrt{\zeta  Q}}+\frac{6 \sqrt{\pi }}{\sqrt{S}}\right), \quad X=-\frac{Q^3 (\log (16 \zeta  Q)-2)}{6 (\zeta  Q)^{3/2}}, \quad T=\frac{8 P S^2-\pi  Q^2+S}{4 \sqrt{\pi } S^{3/2}}.
\end{equation}
The Smarr relation can be expressed as
\begin{equation}\label{Tr}
2TS+\chi Q-2VP+X \zeta=M +\frac{Q^3 \zeta\log (2-16 \zeta  Q)}{6 (\zeta  Q)^{3/2}}=\mathcal{M}.
\end{equation}
where $\mathcal{M}$ is the effective mass. In the absence of $\zeta$, we can obtain
\begin{equation}\label{Tr}
2TS+\chi Q-2VP=M.
\end{equation}
The first law of the BH thermodynamics and the corresponding Smarr relation give a consistent framework for examining a thermodynamic behaviour of the considered charged AdS black hole which establish a relationship among the mass, temperature, entropy, electric charge, and cosmological pressure of the black hole and for the logarithmic correction due to the electric charge of the black hole. For examining the critical behavior of BH, we analyze the thermodynamic topology, introduced in \cite{51}, based on Duan's \(\Phi\) mapping topological theory, which provides a systematic way to identify and classify critical points. The BH temperature can be formulated as a function of its entropy \(S\), pressure \(P\), and other thermodynamic parameters \(x_{i}\) which is defined as \(T = T (S, P, x_{i}).\) Eliminating the state parameter using \((\partial_{S} T)_{P, x_{i}} = 0\), we define Duan's potential \(\Phi\) as
\begin{equation}\label{4}
\Phi = (\sin\theta)^{-1} T(S, x_{i}).
\end{equation}
In this formalism, the factor \((\sin\theta)^{-1}\) is establishing the connection. According to this theory, the associated two-dimensional vector field is defined as \(\phi = (\phi^{S}, \phi^{\theta})\), with components given by
\begin{equation}
\phi^{\theta} = \left(\frac{\partial \Phi}{\partial
\theta}\right)_{S, x_{i}}, \quad \phi^{S} = \left(\frac{\partial
\Phi}{\partial S}\right)_{\theta, x_{i}}.
\end{equation}
The fact is that, this vector field \(\Phi\) has a fixed zero point at \(\theta = \pi/2\), which is used to identify the critical points. Moreover, the conservation of topological charge \(j^{u}\) satisfy \(\phi^a(x_i)=0\), which confirms the existence of some topological current \(W\) is computed via
\begin{equation}
W = \int j^{0} \, d^{2}x = \sum_{i=1}^{N} w_{i} = \sum_{i=1}^{N}
\beta_{i} n_{i},
\end{equation}
where \(j^{0}\) is the topological current density, \(w_{i}\) is the winding number of the \(i\)-th zero point of \(\phi\), \(\beta_{i}\) is the Hopf index, and \(n_{i}\) is the Brouwer degree. In addition to Duan’s \(\Phi\)-mapping approach, an alternative method studies BH models within incomplete thermodynamic parameter spaces. This perspective analyzes the deficiency of winding numbers to probe thermal stability, where the total winding number is identified as the topological charge \cite{52}. Within this framework, the free energy $F$ is defined as
\begin{equation}\label{F0}
F=E\;-\;S \tau^{-1},
\end{equation}
where \(E\) is potential energy, \(S\) is entropy and \(\tau\) is a time-like quantity. The associated vector field  \(\phi = (\phi^{S}, \phi^{\theta})\) is
\begin{equation}\label{5}
\phi = \left( \frac{\partial F}{\partial r_{+}}, -\cot\theta \,
\csc\theta \right),
\end{equation}
which has a zero point at \(\theta = \pi/2\). The unit vector is
defined by
\begin{equation}
n^{a} = \frac{\phi^{a}}{\|\phi\|}, \quad \phi^{1} = \phi_{r}, \quad
\phi^{2} = \phi_{\theta},
\end{equation}
with \(a = 1, 2\). The winding numbers need to be calculated at each zero of the first component of the normalized vector field \(n^a\) (i.e., the zero point of \(n^1\), when the control parameter \(\tau\) is fixed). The winding number is obtained by mapping the thermodynamic behavior of the system, such as free energy, temperature, or horizon radius into a topological space, where it determine the number of thermodynamic defects or distinct BH phases. The zero winding number implies to the absence of a stable phase, whereas the winding number of one signifies the existence of a single stable BH phase as discussed in \cite{52}. Thus, the topological class is defined as follows:  
\begin{equation}\label{6}
W = (2 \pi )^{- 1} \Omega = \frac{1}{2\pi} \oint_{C} \epsilon_{ab} n^{a} \partial_{i} n^{b} d x^{i}.
\end{equation}
Here, \(n^a\) is the unit vector field constructed from thermodynamic potential, \(\epsilon_{ab}\) is the antisymmetric tensor, and the integration is performed along a closed contour $C$ in the thermodynamic parameter space. Here, a family of closed curves parameterized by $\vartheta \in (0, 2\pi)$ as
\begin{equation}\label{7}
\left\{
\begin{aligned}
r &= a \cos\vartheta + r_0,\\
\theta &= b \sin\vartheta + \frac{\pi}{2}.
\end{aligned}
\right.
\end{equation}
After choosing \((a,b,r_{0})\) for the closed curves to encircle candidate defects, we evaluate the field (or a related quantity derived from the free energy) along these contours and compute the line integral. The sign of each contribution reflects the local rotation (counter-clockwise  being +ve and clockwise -ve), while the total sum \(W\) representing the combined contribution of all enclosed zeros yields the conserved topological charge characterizing the phase structure; a change in \(W\) signifies the topological transition associated with thermodynamic phase transitions. We select \((a, b, r_{0})\), subsequently calculate \(\mathcal{F}\) along these contours. Within the standard Boltzmann-Gibbs framework, the entropy of a BH is given by the Bekenstein-Hawking area law, $S_0 = \frac{A_{\text{BH}}}{4}, \quad A_{\text{BH}} = 4\pi r_h^2,$ where $A_{\text{BH}}$ represents the area of the event horizon \cite{Ref3}. This formulation relies on the assumptions of extensivity and additivity, however, these assumptions may break down in the presence of long range interactions, the strong gravitational fields, or quantum effects. To address such non-extensive effects, generalized entropy formalisms, including Tsallis entropy and R\'{e}nyi entropy, have been proposed. Sharma–Mittal entropy \cite{53,54,55} offers a more general formalism, defined as follows:
\begin{equation}\label{S0}
S = \frac{1}{\gamma} \left[ \left( \alpha
\frac{S_{\mathrm{BH}}}{\beta} + 1 \right)^{\beta} - 1 \right],
\end{equation}
where $\alpha$ and $\beta$ are entropy constants satisfying $\alpha, \beta > 0$, with \(\alpha \neq 1 \neq \beta\), \(\alpha \neq \beta\), and \(0 < \gamma < 1\). The advantage of Sharma-Mittal entropy is its ability to smoothly interpolate between different entropy measures, particularly significant for BHs, whose thermodynamics may deviate from standard extensivity bacause of event horizon and strong gravitational field. Another advantage of the Sharma–Mittal entropy is that it can effectively characterize the non-additivity and non-extensivity that are naturally present in self-gravitating systems. BHs are dominated by long-range gravitational interactions, and the Sharma–Mittal can describe such systems more realistically than the conventional entropy definitions. Therefore, it offers deeper insights into understanding of thermodynamic stability, critical behavior, and phase transitions.


\section{Canonical Ensemble}\label{B}
\begin{figure}[h!]
\centering
\includegraphics[width=8cm]{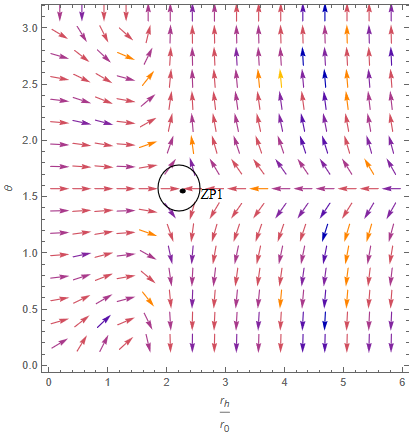}
\caption {\scriptsize Normalized vector field $n$ plotted in the
$(r_h/r_0,\theta)$ plane for the BH system. The dot
indicates the critical point.} \label{fig.N}
\end{figure}

To examine the thermodynamic topology of BH with generalized entropy for considering canonical ensemble, the mass \(M\) derived from Eq. (\ref{1}), Hawking temperature \(T_H\) of charged AdS black holes can be written in terms of horizon radius \(r_h\) as 
\begin{equation}\label{M}
M=\frac{1}{6} \left( \frac{3 Q^2}{r_h} + 3{r_h} + 8 P \pi {r_h}^3 + \frac{2
Q^2 \ln(16 Q \zeta)}{\sqrt{Q \zeta}} \right).
\end{equation}
It is necessary to express Hawking temperature as a function of the pressure and charge as
\begin{equation}\label{TH}
T_H = \frac{\partial M}{\partial S}= \frac{ \gamma \left( {r_h^2} -
Q^2 + 8 P \pi r_h^4 \right)\left( 1 + \frac{\pi r_h^2 \alpha}{\beta} \right)^{1 - \beta} }{ 4 \pi r_h^3 \alpha }.
\end{equation}
In extended black hole thermodynamics, the critical pressure is obtained by the inflection point required by the Hawking temperature T with respect to the event horizon radius rh, denotes the presence of second order phase transition which satisfies the condition $\left( {\partial T}/{\partial r_h} \right)_{Q} = 0$, $\left( {\partial^2 T}/{\partial r_h^2} \right)_{Q} = 0$. Calculating both the equations simultaneously that provides the critical horizon radius and corresponding pressure as discussed in Ref \cite{Ref1}.

\begin{equation}\label{P}
P = \frac{ \left( (3 Q^2 - r_h^2) \beta + \pi r_h^2 \alpha \left( r_h^2 (1
- 2\beta) + Q^2 (1 + 2\beta) \right) \right) }{ 8 \pi r_h^4 \left(
-\beta + \pi r_h^2 \alpha (-3 + 2\beta) \right) }.
\end{equation}
Substituting this result into the temperature relationship in Eq. (\ref{TH}) yields the corresponding isobaric temperature $T$ becomes
\begin{equation}\label{TP}
T_P=\frac{\gamma\; (2 Q^2 - r_h^2) (\pi r_h^2 \alpha + \beta)^2\left(1 + \frac{\pi r_h^2 \alpha}{\beta}
\right)^{-\beta} }{ 2 \pi r_h^3
\alpha \beta \left( -\beta + \pi r_h^2 \alpha (-3 + 2 \beta) \right)}.
\end{equation}
In accordance with the topological method outlined in the introduction, the parameter space is subsequently examined to locate the zero points of the associated vector field. These zero points correspond to the on-shell BH solutions, whose topological characteristics are encoded in the corresponding charge. The associated topological charge is then determined by the $\Phi$-mapping formalism with
topological current theory. The thermodynamic function \(\Phi = \frac{1}{\sin \theta} \, T ,\) is defined in Eq. (\ref{4}) as
\begin{equation}
\Phi =\frac{ (2 Q^2 - r_h^2) \left( 1 + \frac{\pi r_h^2 \alpha}{\beta}
\right)^{-\beta} (\pi r_h^2 \alpha + \beta)^2 \gamma \csc\theta }{ 2
\pi r_h^3 \alpha \beta \left( -\beta + \pi r_h^2 \alpha (-3 + 2 \beta)
\right) }.
\end{equation}
 The vector field's components  \(\phi = (\phi^{r_h}, \phi^{\theta})\) are
\begin{equation}\label{phi1}
\phi^{r_h} ={\frac{ \left( 1 + \frac{\pi r_h^2 \alpha}{\beta}
\right)^{-\beta} (\pi r_h^2 \alpha + \beta) \;A\; \gamma \csc\theta }{
2 \pi r_h^4 \alpha \beta \left( \pi r_h^2 \alpha (3 - 2\beta) + \beta
\right)^2 }},
\end{equation}
where $A$ is defined in Appendix in Eq. (\ref{A}) and
\begin{equation}
\phi^\theta = \frac{ (2 Q^2 - r_h^2) \left( 1 + \frac{\pi r_h^2
\alpha}{\beta} \right)^{-\beta} (\pi r_h^2 \alpha + \beta)^2 \gamma
\cot\theta \csc\theta }{ 2 \pi r_h^3 \alpha \beta \left( \beta + \pi
r_h^2 \alpha (3 - 2\beta) \right) }.
\end{equation}
The normalized vector's component is defined as $\vec{n} = \left(
{\phi_{r_h}}/{\| \phi \|}, {\phi_\theta}/{\| \phi \|} \right)$ are expressed as
\begin{equation}\label{8}
\frac{\phi^{r_h}}{\left\| \phi \right\|} ={\frac{ A }{
\left[{ {A}^2 + \left(r_h (-2 Q^2 + r_h^2) (\pi r_h^2 \alpha + \beta) (\pi
r_h^2 \alpha (3 - 2 \beta) + \beta) \cot\theta
\right)^2}\right]^\frac{1}{2}}},
\end{equation}
and
\begin{equation}\label{9}
\frac{\phi^\theta}{\left\| \phi \right\|}=\frac{ - r_h \left( \pi r_h^2 \alpha (3 - 2\beta) + \beta \right) (-2 Q^2 + r_h^2) \cot\theta}{ \left[{A}^2 + \left(r_h (-2 Q^2 + r_h^2) (\pi r_h^2 \alpha + \beta) (\pi r_h^2 \alpha (3 - 2 \beta) + \beta) \cot\theta \right)^2 \right]^\frac{1}{2}}.
\end{equation}
The behavior of normalized vectors in the \( r_h \) vs \( \theta \) plane for the BH is illustrated in Fig. \ref{fig.N}. The plot is generated using the parameter values  \( \alpha=1.1\), \(\beta=1.2\), \(\gamma=0.1\) and \( Q = 1.01 \). The black dot in the figure marks the critical point, denoted by $ZP1$. To determine this critical point, one sets \( \theta = \tfrac{\pi}{2} \) in Eq.~\eqref{8} for zero pint $ZP1$ and solves the condition under which the expression vanishes.

\begin{figure}[htbp]
    \centering
    \begin{subfigure}[t]{0.49\textwidth}
        \centering
        \includegraphics[width=8cm]{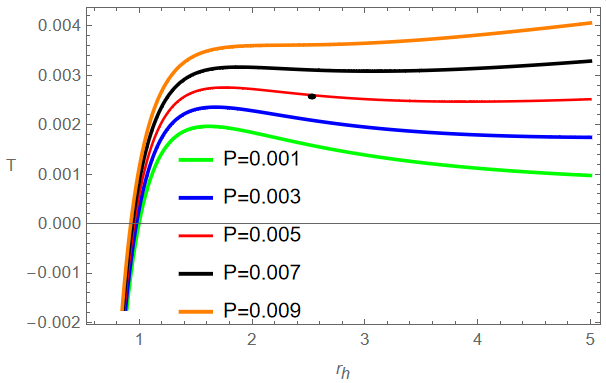}
        \label{fig:T1}
    \end{subfigure}
    \hfill
    \begin{subfigure}[t]{0.49\textwidth}
        \centering
        \includegraphics[width=8cm]{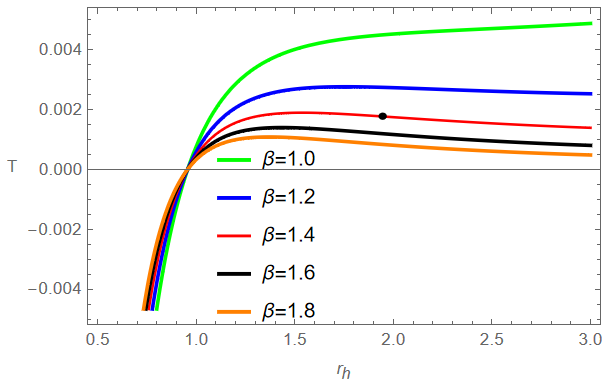}
        \label{fig:beta1}
    \end{subfigure}
    \caption{\scriptsize Variation of the isobaric temperature \(T\) with the horizon radius \(r_h\)
    for a BH at different values of the parameter \(P\) and $\beta$). The critical point
    is marked by the dot. }\label{TB1}
\end{figure}
The graphical behavior shows that the temperature curve can intersect the zero axis at various points depending on the parameter choices, producing multiple zero points within the thermodynamics phase space. The Black dot indicates the critical point in the phase structure in Fig. \ref{TB1}. The first graph demonstrate the effects of the thermodynamic pressure \(P\) on the temperature profile while the other graph indicates the dependence of the isobaric temperature on the parameter \(\beta\) . The red curve indicates the isobaric curve at \(P=P_c\), while the above and below curves represents the \(P>P_c\) and  \(P<P_c\) region, respectively. At lower values pressure and \(\beta\) , the temperature initially rises, reaches a maximum, and then decreases below zero, clearly crossing the zero axis.

\begin{figure}[htpb]
\centering
\includegraphics[width=8cm]{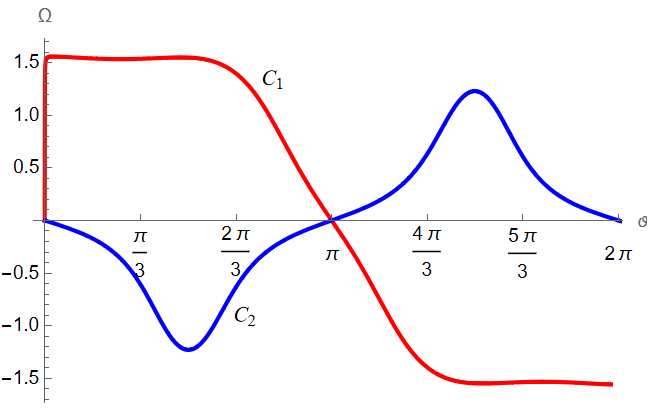}
\caption {\scriptsize Variation of $\Omega$ w.r.t. $\vartheta$ along the contours $C_1$ and $C_2$, showing the angular dependence of $\Omega$ across different
trajectories}\label{O1}
\end{figure}

Fig. \ref{O1} shows the dependence of $\Omega(\vartheta)$ on the
angular parameter $\vartheta \in [0,2\pi]$ with \((a, b, r_{0}) = (0.237, 1.50, 2.038)\) for \(C_{1}\) and \((a, b, r_{0}) = (0.1, 1.5, 1.0)\) for \(C_{2}\). The red curve represents \(C_{1}\) displays a monotonic behavior of $\Omega(\vartheta)$, beginning from a positive value at $\vartheta=0$, decreasing smoothly, crossing zero at $\vartheta=\pi$, and approaching a negative constant as $\vartheta \to 2\pi$. The total variation across one full cycle yields a winding number of $-1$. While, the blue curve represents \(C_{2}\), an oscillatory solution of $\Omega(\vartheta)$, which vanishes at the boundaries $\vartheta=0,\pi,2\pi$. Hence, for the contour $C_{1}$, the topological charge is found to be ${\omega_{1}}=-1$, while for the contour $C_{2}$, the topological charge vanishes, ${\omega_{2}}=0$. Thus, the total topological charge is entirely determined by the red curve, giving a net value $W=-1$. The corresponding critical points are determined accordingly.
\begin{equation}\label{cp}
r_c= \sqrt{2} \, Q,\\
\quad P_c = \frac{-\beta -2 \pi  \alpha  (10 \beta -7) Q^2}{96 \pi  Q^2 \left(6 \pi  \alpha  (2 \beta -3) Q^2-\beta \right)},\\
\quad T_c=\frac{\gamma \left(2 \pi  \alpha  (2 \beta -5) Q^2-\beta \right)}{2 \sqrt{2} \pi  \alpha  Q \left(18 \pi  \alpha  (2 \beta -3) Q^2-3 \beta \right)} \left(\frac{2 \pi  \alpha  Q^2}{\beta }+1\right)^{1-\beta }.
\end{equation}

\section{Canonical Ensemble as BH Solutions}\label{C}
Now, we proceed to study in canonical ensemble as topological thermodynamic defects. Substituting the BH mass and entropy into the general expression for the generalized free energy renders this quantity especially useful for exploring thermodynamic properties, such as phase transitions and critical behavior. Using the mass from Eq. (\ref{M}) together with the entropy of the BH from Eq. (\ref{S0}) in  Eq. (\ref{F0}), the generalized free energy is obtained as:

\begin{equation}
F=\frac{1}{6} \left( \frac{3 Q^2}{r_h} + 3r_h + 8 P \pi r_h^3 - \frac{6 \left( -1 + \left(1 + \frac{\pi r_h^2 \alpha}{\beta} \right)^{\beta}
\right)}{\gamma \tau} + \frac{2 Q^2 \ln (16 Q \zeta)}{\sqrt{Q
\zeta}} \right).
\end{equation}
The components of vector field are calculated as:
\begin{equation}
\phi^{r_h} = \frac{1}{2} - \frac{Q^2}{2 r_h^2} + 4 P \pi r_h^2 - \frac{2 \pi
r_h \alpha \left( 1 + \frac{\pi r_h^2 \alpha}{\beta} \right)^{-1 +
\beta}}{\gamma \tau}.
\end{equation}
and $\phi^\theta$ gives in Eq. (\ref{5}). The normalized vectors are expressed as:
\begin{equation}
\frac{\phi^{r_h}}{\left\| \phi \right\|} = \frac{ \frac{1}{2} -
\frac{Q^2}{2 r_h^2} + 4 P \pi r_h^2 - \frac{2 \pi r_h \alpha \left( 1 +
\frac{\pi r_h^2 \alpha}{\beta} \right)^{-1 + \beta}}{\gamma \tau} }{
\left[{ \left( \frac{1}{2} - \frac{Q^2}{2 r_h^2} + 4 P \pi r_h^2 -
\frac{2 \pi r_h \alpha \left( 1 + \frac{\pi r_h^2 \alpha}{\beta}
\right)^{-1 + \beta}}{\gamma \tau} \right)^2 + \cot^2\theta \,
\csc^2\theta } \right]^\frac{1}{2}},
\end{equation}
and
\begin{equation}
\frac{\phi ^\theta}{\left\| \phi \right\|} = \frac{ - \cot\theta
\csc\theta }{ \left[{ \left( \frac{1}{2} - \frac{Q^2}{2 r_h^2} + 4 P
\pi r_h^2 - \frac{2 \pi r_h \alpha \left( 1 + \frac{\pi r_h^2
\alpha}{\beta} \right)^{-1 + \beta}}{\gamma \tau} \right)^2 +
\cot^2\theta \, \csc^2\theta } \right]^\frac{1}{2}}.
\end{equation}
\begin{figure}[h!]
    \centering
    \begin{subfigure}[b]{0.48\textwidth}
        \centering
        \includegraphics[width=\textwidth]{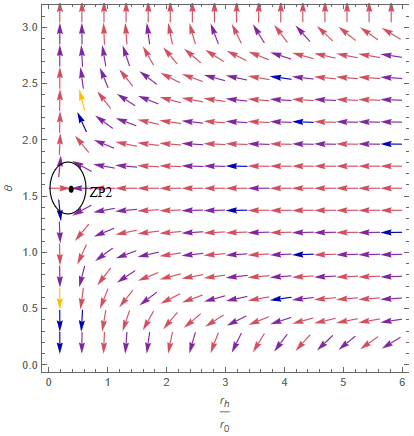}\label{Vec2}
    \end{subfigure}
    \hfill
    \begin{subfigure}[b]{0.48\textwidth}
        \centering
        \includegraphics[width=\textwidth]{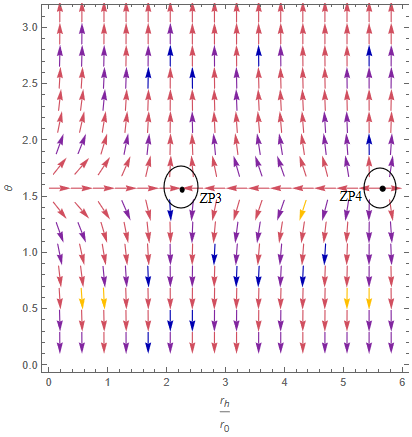}\label{Vec3}
    \end{subfigure}
    \caption{\scriptsize Normalized vector field $n$ plotted in the $(r_h/r_0,\theta)$ plane for the BH system using $\tau=40$ for $ZP2$ in left panel and  $\tau=250$ for $ZP3$ and $ZP4$ in right panel. The dot indicates the critical point}\label{Vec2}
\end{figure}
The zero points are determined by setting the angle \( \theta = {\pi}/{2} \) in \( \frac{\phi_{r_h}}{\left\| \phi \right\|} \) and imposing the condition that the expression equals zero. As an illustration, for chosen parameter values: \(\alpha = 1.05 \), \(\beta = 1.5 \), \(\gamma = 0.1 \) and using the ratios \(\frac{Q}{r_0} = 0.1 \), \( P r_0^2 = 0.005 \), and \(\frac{\tau}{r_0} = 40\), we obtain a zero point \( \text{ZP2} \), as shown in left side of Fig. \ref{Vec2} at the coordinates $\left( {r_h}/{r_0}, \Theta \right) = (0.39, \frac{\pi}{2}).$ The pressure in this case is considered to be below the critical pressure \( P_c \). Likewise, using the parameters \(\alpha = 1.01\), \(\beta = 1.125\), \(\gamma = 0.1\), \(\frac{Q}{r_0} = 0.1\), \(P r_0^2 = 0.005\), with \(\frac{\tau}{r_0} = 250 \), we find two distinct zero points, \(\text{ZP3} \) and \( \text{ZP4} \), as demonstrated in right side of Fig. \ref{Vec2}. The associated topological numbers at these critical points may change, signaling possible transitions between distinct thermodynamic phases. By systematically investigating different combinations of \(\alpha\), \(\beta\), and \(\gamma\), thus determines whether a phase transition persists, strengthens, weakens, or disappears.\\

\begin{figure}[h!]
\centering
\includegraphics[width=8cm]{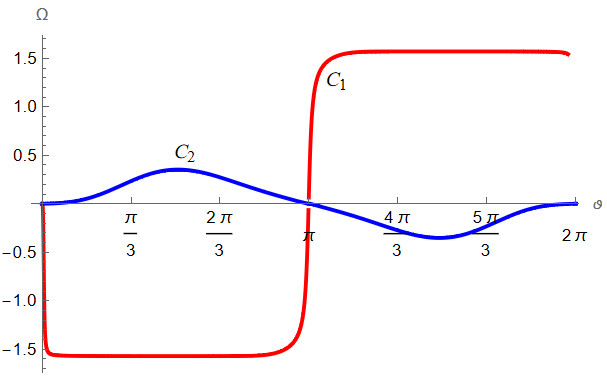}
\caption {\scriptsize Variation of $\Omega$ w.r.t. $\vartheta$ along the contours $C_1$ and $C_2$, showing the angular dependence of $\Omega$ across different trajectories}
\label{O2}
\end{figure}
Figure \ref{O2} illustrates the topological charge with the total variation across the interval yields a winding number of $+1$. For the contour $C_{1}$ associated with the red curve with \((a, b, r_{0}) = (0.9, 1.55, 1.9)\), the topological charge is ${\omega_{1}}=+1$, while for $C_{2}$ associated with the blue curve with \((a, b, r_{0}) = (1.5, 0.35, 4.309)\), the charge vanishes. The net topological charge is thus determined entirely by the red curve, yielding $W=+1$. By imposing the condition \( \phi^{r_h} = 0 \), one can derive an analytical expression for \( \tau \) corresponding to the zero point

\begin{equation}
\tau = \frac{ 4 \pi r^3 \alpha \left( 1 + \frac{\pi r^2
\alpha}{\beta} \right)^{\beta} \beta }{ (-Q^2 + r^2 + 8 P \pi
r^4)(\pi r^2 \alpha + \beta) \gamma }.
\end{equation}

The variation of the generalized parameters $\alpha, \beta > 0$, with \(\alpha \neq 1 \neq \beta\), \(\alpha \neq \beta\), and \(0 < \gamma < 1\), plays a significant role in BH thermodynamics. Since \(\alpha\), \(\beta\), and \(\gamma\) appear in the expressions for the \(\tau\)-function and the vector field components (\(n_1, n_2\)), their values directly determine the positions of critical points and shape the overall phase structure of the BH.

\section{Mixed Ensembles}\label{D}
In the mixed ensemble of BH thermodynamics, the electric potential \( \phi_e \) kept constant which is defined as $\phi_e = {Q}/{r_h}.$ The modified mass according to the above assumption is
\begin{equation}
M = \frac{1}{6} r_h \left( 3 + 8 P \pi r_h^2 + 3 \phi_e^2 + \frac{2
\phi_e \sqrt{r_h \zeta \phi_e} \ln (16 r_h \zeta \phi_e)}{\zeta}
\right),
\end{equation}
and the modified temperature is
\begin{equation}
T = \frac{ \left(1 + \frac{\pi r_h^2 \alpha}{\beta} \right)^{1 -
\beta} \, \gamma \left( 2 \phi_e \sqrt{r_h \zeta \phi_e} + 3 \zeta
\left(1 + 8 P \pi r_h^2 + \phi_e^2\right) + 3 \phi_e \sqrt{r_h \zeta
\phi_e} \ln (16 r_h \zeta \phi_e) \right) }{12 \pi r_h \alpha \zeta}.
\end{equation}
Thermodynamics function is $\Phi = \frac{T}{\sin\theta}$, and the vector field \( {\phi} = (\phi^{r_h}, \phi^{\theta}) \) has the following vector components which are written as
\begin{equation}\label{phi2}
\phi^{r_h} = -\frac{1}{24 \pi r_h^2 \alpha \beta \sqrt{r_h \zeta
\phi_e}} \left(1 + \frac{\pi r_h^2 \alpha}{\beta} \right)^{-\beta}
\gamma \, \csc\theta \left( 2 \Big( 24 P \pi^2 r_h^4 \alpha (-3
+ 2\beta) \sqrt{r_h \zeta \phi_e} + \;B\; \Big) + 3 r_h\; C\; \phi_e^2
\ln (16 r_h \zeta \phi_e) \right),
\end{equation}
 where $B$ and $C$ are defined in Appendix in Eq.s (\ref{b}) and (\ref{c}), and
\begin{equation}
\phi^\theta = -\frac{\left(1 + \frac{\pi r_h^2 \alpha}{\beta}
\right)^{1 - \beta} \, \gamma \, \cot\theta \, \csc\theta \left( 2
\phi_e \sqrt{r_h \zeta \phi_e} + 3 \zeta (1 + 8 P \pi r_h^2 + \phi_e^2)
+ 3 \phi_e \sqrt{r_h \zeta \phi_e} \ln (16 r_h \zeta \phi_e) \right)}{12
\pi r_h \alpha \zeta}.
\end{equation}
The components of the normalized vector \(\phi\) are derived as
\begin{equation}\label{N1}
n^1=\frac{{\sqrt\zeta} \csc\theta \, \Bigg( 2 \Big( 24 P \pi^2 r_h^4
\alpha (-3 + 2\beta) \sqrt{r_h \zeta \phi_e} +\; B\; \Big) + 3r_h\; C\;
\phi_e^2 \ln\!\big(16r_h\zeta\phi_e\big)\Bigg)}{\left[ \csc^2\theta
\Bigg( \zeta \Big( 2\Big( 24 P \pi^2 r_h^4 \alpha (-3+2\beta)\sqrt{r_h
\zeta \phi_e} + \;B\; \Big) + 3r_h\;C\;\phi_e^2 \ln(16 r_h \zeta \phi_e)
\Big)^2 + 4 r_h^3 (\pi r_h^2 \alpha + \beta)^2 \phi_e \cot^2\theta D ^2
\Bigg)\right]^\frac{1}{2} },
\end{equation}
where $D$ also defined in Appendix in Eq. (\ref{d}), and
\begin{equation}
n^2 = -\,\frac{2r_h^{\frac{3}{2}}\beta\sqrt{\phi_e}\left(1+\frac{\pi
r_h^{2}\alpha}{\beta}\right)\cot\theta\,\csc\theta\,
\Big(2\phi_e\sqrt{r_h\zeta\phi_e}+3\zeta\big(1+8P\pi
r_h^{2}+\phi_e^{2}\big)+3\phi_e\sqrt{r_h\zeta\phi_e}\,\ln \!\big(16r_h\zeta\phi_e\big)\Big)}
{\left[{\csc^{2}\theta\,\Bigg(\zeta\Big(2\big(24P\pi^{2}r_h^{4}\alpha(-3+2\beta)\sqrt{r_h\zeta\phi_e}
+\;B\;\big)+3r_h\;C\;\phi_e^{2}\ln \!\big(16r_h\zeta\phi_e\big)\Big)\Bigg)^{2}
+4r_h^{3}\big(\pi r_h^{2}\alpha+\beta\big)^{2}\phi_e\cot^{2}\theta\,
D^{2}}\right]^\frac{1}{2}}.
\end{equation}

\begin{figure}[h!]
    \centering
    \begin{subfigure}[b]{0.48\textwidth}
        \centering
        \includegraphics[width=\textwidth]{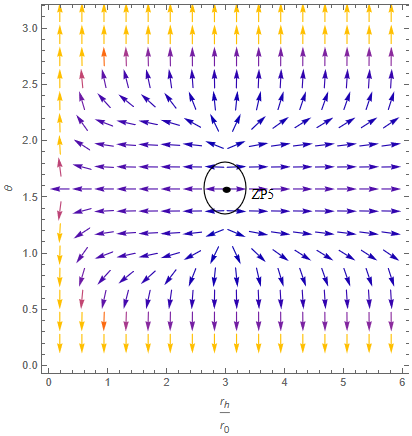}
        \subcaption{\scriptsize  Normalized vector field $n$ plotted in the $(r_h/r_0,\theta)$ plane for the BH system. The dot indicates the critical point.}\label{Vec4}
    \end{subfigure}
    \hfill
    \begin{subfigure}[b]{0.48\textwidth}
        \centering
        \includegraphics[width=\textwidth]{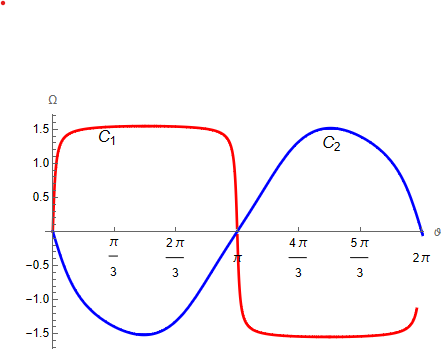}
        \subcaption{\scriptsize Variation of $\Omega$ with respect to $\vartheta$ along the contours $C_1$ and $C_2$, illustrating the angular dependence of $\Omega$ across different trajectories.}\label{O3}
    \end{subfigure}
\end{figure}
The zero point is determined by setting the angle \( \theta = \frac{\pi}{2} \) in \( \frac{\phi_r}{\left\| \phi \right\|} \) and imposing the condition that the expression equals zero. As an illustration, for chosen parameter values: \(\alpha = 1.5 \), \(\beta = 1.07 \), \(\gamma = 0.1 \) and using the ratios \(\frac{Q}{r_0} = 0.1 \), \( P r_0^2 = 0.005 \), and \(\zeta = 0.1 \), we obtain a zero point \( \text{ZP5} \), as shown in Fig. \ref{Vec4} at the coordinates $\left( {r_h}/{r_0}, \Theta \right) = (3.01,\frac{\pi}{2}).$ Figure \ref{O3} illustrates the topological charge with the total variation across the interval yields a winding number of $+1$. For the contour $C_{1}$ associated with the red curve with \((a, b, r_{0}) = (0.5,1.0,8.99)\), the topological charge is ${\omega_{1}}=-1$, while for $C_{2}$ associated with the blue curve with \((a, b, r_{0}) = (0.2,1.5,0.9)\), the charge vanishes. The net topological charge is thus determined entirely by the red curve, yielding $W=-1$. 

\begin{figure}[h!]
    \centering
    \begin{subfigure}[t]{0.49\textwidth}
\centering
\includegraphics[width=8cm]{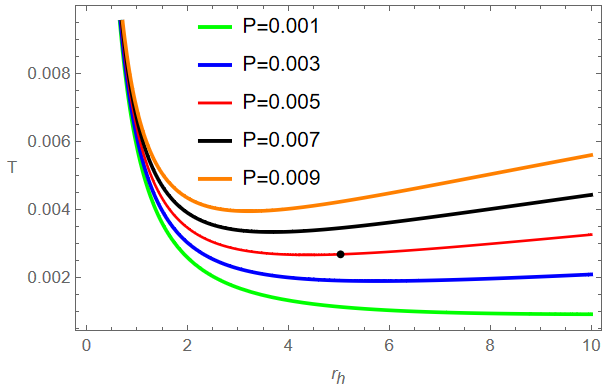}\label{T2}
    \end{subfigure}%
    \hfill
    \begin{subfigure}[t]{0.49\textwidth}
\centering
\includegraphics[width=8cm]{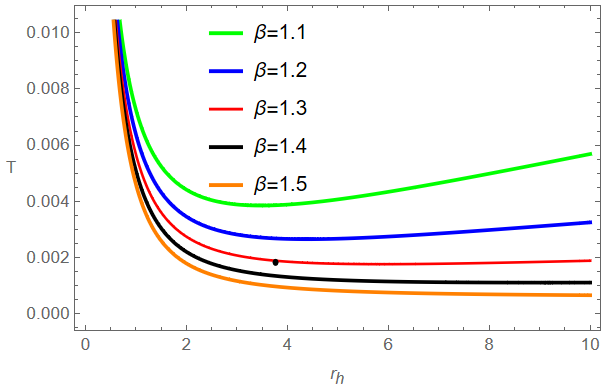}\label{beta2}
\end{subfigure}
 \caption{ Variation of the isobaric temperature \(T\) with the horizon radius \(r_h\)
for a BH at different values of the parameter \(P\) in left side and $\beta$ in right side. The critical point is
marked by the dot. }\label{TB0}
\end{figure}
In Fig. \ref{TB0}, the temperature is plotted for different values of pressures \(P\). Here, the curves rise with increasing pressure, the green curve lies lowest. For lower pressures (\(P < P_c\)), each curve shows two extremal points, corresponding to the separation between stable small and large BH phases, while for higher pressures (\(P > P_c\)) the curves become monotonic, indicating the disappearance of the first-order phase transition. Also the temperature profile is shown for different values of \(\beta\). The green curve exhibits a sharp minimum, while as \(\beta\) increases in other curves, the curves gradually flatten, shifting the extremal points and reducing the temperature at larger horizon radii, as shown in Fig. \ref{TB0}. The black dot indicates the critical point in phase structure.

\section{Topological Thermodynamic Flaws in Mixed Ensemble}\label{E}
We next analyze the mixed ensemble from the
perspective of a topological defect. As a starting point, we
consider the Gibbs free energy potential, where $Q$ represents
the electric charge in the BH framework, (defined as $F = M - {\phi_e} Q - {S}/{\tau}$). In this ensemble $Q = {r_h}{\phi_e}$, the equation becomes
\begin{equation}
F= \frac{1}{6} r_h \left(3 + 8 P \pi r_h^2 + 3 \phi_e^2 + \frac{2 \phi_e \sqrt{r_h \zeta \phi_e} \ln (16 r_h \zeta \phi_e)}{\zeta} \right) - r_h {\phi_e}^2 - \frac{-1 + \left(1 + \frac{\pi r_h^2 \alpha}{\beta}\right)^{\beta}}{\gamma \tau},
\end{equation}
this can be written in the form of potentials as
\begin{equation}\label{E1}
\phi^{r_h} = \frac{1}{6} \left( 3 + 24 P \pi r_h^2 + \frac{3 \phi_e\sqrt{r_h \zeta \phi_e} \ln (16 r_h \zeta \phi_e)}{\zeta} +\;E\; \right).
\end{equation}
where $E$ defined in Appendix in Eq. (\ref{e}), and $\phi^\theta$ given in Eq. (\ref{5}). The corresponding normalized vectors are as
\begin{equation}
\frac{\phi^{r_h}}{\left\| \phi \right\|} = \frac{ \frac{1}{6} \left( 3 +
24 P \pi r_h^2 + \frac{3 \phi_e \sqrt{r_h \zeta \phi_e} \ln (16 r_h \zeta \phi_e)}{\zeta} +\;E\; \right) } { \left[{ \left( \frac{1}{6} \left( 3 + 24 P \pi r_h^2 + \frac{3 \phi_e \sqrt{r_h \zeta \phi_e}
\ln (16 r_h \zeta \phi_e)}{\zeta} +\;E\; \right) \right)^2 + \cot^2 \theta
\csc^2 \theta }\right]^\frac{1}{2} },
\end{equation}
and
\begin{equation}
\frac{\phi^\theta}{\left\| \phi \right\|} = \frac{- \cot \theta \csc
\theta} {\left[{ \left( \frac{1}{6} \left( 3 + 24 P \pi r_h^2 +
\frac{3 \phi_e \sqrt{r_h \zeta \phi_e} \ln (16 r_h \zeta \phi_e)}{\zeta} +\;E\;
\right) \right)^2 + \cot^2 \theta \csc^2 \theta.
}\right]^\frac{1}{2}}.
\end{equation}
The analytic expression for $\tau$ is derived by imposing the
condition $\Phi^{r_h} =0$, which yields
\begin{equation}
\tau = \frac{ 12 \pi r_h  \alpha  \beta  \zeta \left(1 + \frac{\pi r_h^2 \alpha}{\beta}
\right)^\beta }{ \gamma \left( \pi r_h^2 \alpha + \beta \right)  \left( 2 \phi_e \sqrt{r_h \zeta \phi_e} + 3 \zeta (1 + 8 P \pi
r_h^2 - \phi_e^2) + 3 \phi_e \sqrt{r_h \zeta \phi_e} \ln (16 r_h \zeta
\phi_e) \right) }.
\end{equation}

\begin{figure}{h}
    \centering
    \begin{subfigure}[b]{0.48\textwidth}
        \centering
        \includegraphics[width=\textwidth]{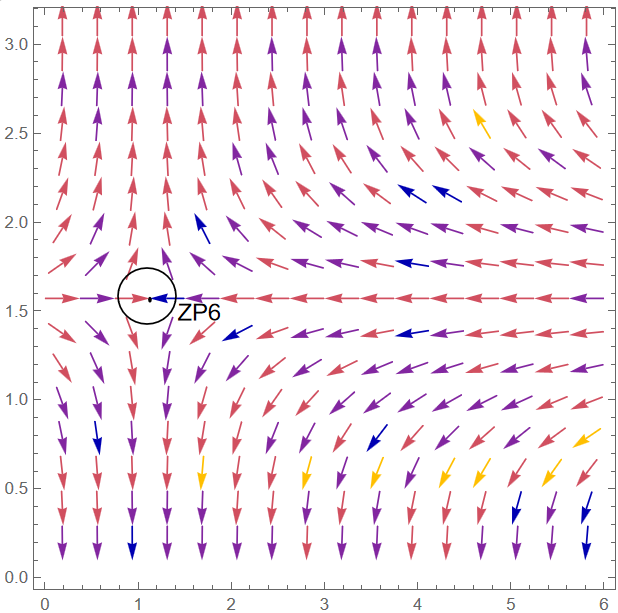}
        \subcaption{\scriptsize  Normalized vector field $n$ plotted in the $(r_h/r_0,\theta)$ plane for the BH system. The dot indicates the critical point.}\label{Vec5}
    \end{subfigure}
    \hfill
    \begin{subfigure}[b]{0.48\textwidth}
        \centering
        \includegraphics[width=\textwidth]{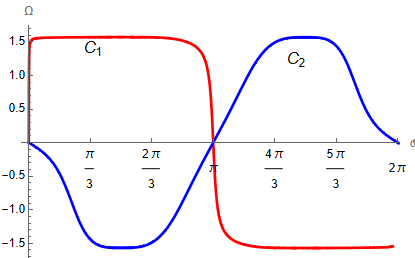}
        \subcaption{\scriptsize Variation of $\Omega$ with respect to $\vartheta$ along the contours $C_1$ and $C_2$, illustrating the angular dependence of $\Omega$ across different trajectories.}\label{O4}
    \end{subfigure}
\end{figure}

The zero point is determined by setting the angle \( \theta = \frac{\pi}{2} \) in \( \frac{\Phi_r}{\left\| \Phi \right\|} \) and imposing the condition that the expression equals zero. As an illustration, for chosen parameter values: \(\alpha = 1.20 \), \(\beta = 1.102 \), \(\gamma = 0.854 \) and using the ratios \( \frac{Q}{r_0} = 0.9 \), \( P r_0^2 = 0.005 \), and \(\zeta = 1.8 \), \(\tau=10\), we obtain a zero point \( \text{ZP6} \), as shown in Fig. \ref{Vec5} at the coordinates $\left( {r_h}/{r_0}, \Theta \right) = (1.045,\frac{\pi}{2}).$ Similarly as shown in Fig. \ref{O1}, the Fig. \ref{O3} illustrates the topological charge for the parameter sets \((a,b,r_{0}) = (1.2,1.6,5.3)\) for \(C_{1}\) associated with the red curve, the topological charge is $Q_{C_{1}}=-1$ and \((a, b, r_{0}) = (0.9,1.55,1.9)\) for \(C_{2}\) (blue curve), the charge vanishes. Therefore, the overall topological charge is determined entirely by the red curve, giving $Q=-1$.

\section{Heat Capacity}\label{G}
The thermodynamic property of BH can also be determined by analyzing the behavior of heat capacity, which is known by its thermal stability. Particularly, a positive behavior of heat capacity signifies thermal stability while negative value corresponds to instability \cite{40,41,42}. The major significance is the divergence of the heat capacity, which is typically associated with BH phase transitions. The heat capacity is expressed as
\begin{equation}\label{12}
C =\frac{T}{\frac{\partial^2 M}{\partial S^2}}=-\frac{(Q^2 - r_h^2 - 8
\pi P r_h^4)\left( \beta + \pi r_h^2 \alpha (1 - 2 \beta) \right)}{2
(Q^2 + 8 \pi P r_h^4)(\pi r_h^2 \alpha + \beta)},
\end{equation}
where
\begin{equation}\label{13}
\frac{\partial^2 M}{\partial S^2}=\frac{ \left( Q^{2} + 8 P \pi
r_h^{4} \right) \left( 1 + \frac{\pi r_h^{2} \alpha}{\beta}
\right)^{-\beta} \left( \pi r_h^{2} \alpha + \beta \right)^{2} \gamma
}{ 2 \pi r_h^{3} \alpha \beta \left[ \beta + \pi r_h^{2} \alpha \left(
-1 + 2 \beta \right) \right] }.
\end{equation}

The physical and non-physical solutions of a BH can be determined
from the roots of temperature: a positive temperature corresponds to
a physical solution, while a negative value indicates a non-physical
one. Once the thermodynamic behavior is established, it becomes
essential to examine the phase transition and bound points of the
heat capacity using Eq.s (\ref{12}) and (\ref{13}). 
\begin{figure}[h!]
    \centering
\includegraphics[width=8cm]{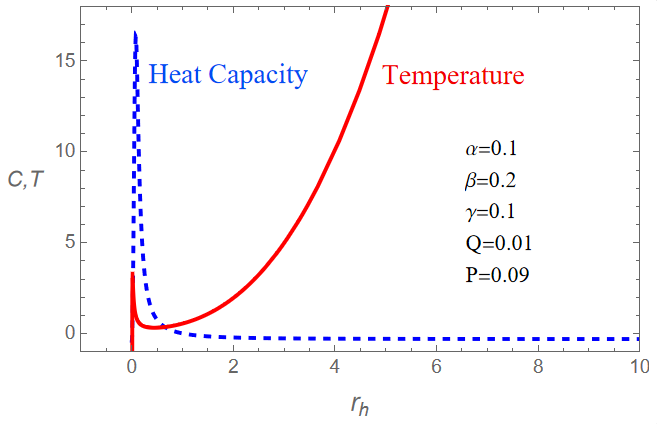}
\caption {\scriptsize The graphical representation illustrates the variation of heat capacity $C$ and a function of temperature $T$.}\label{hc}
\end{figure}
For a BH system to remain physically stable, its heat capacity must be positive. This stability is reflected in the positive behavior of heat capacity, as illustrated in Fig. \ref{hc}, which shows the behavior of temperature and heat capacity with respect to the horizon radius $r_h$. The temperature (red curve) increases monotonically with $r_h$, where the positive branch represents the physical BH solutions. The heat capacity (blue dashed curve) exhibits a divergence at a critical horizon radius, indicating a second-order phase transition. For small $r_h$, the heat capacity is positive but quickly turns negative, reflecting instability. Thus, thermodynamic stability is ensured only in the region where both temperature and heat capacity remain positive.

\section{Particle dynamics}\label{K}
In this section, we examine the motion of particles in the vicinity of the considered BH solution as discussed in Ref. \cite{OME}. For this purpose, we introduce the following Hamiltonian \cite{43,44} as
\begin{equation}
H = \frac{1}{2} g^{\mu \nu} p_{\mu} p_{\nu} + \frac{1}{2} m^{2}. \label{eq:6}
\end{equation}
In the above equation, \( m \) denotes the mass of the particle, and
\( p^{\mu} = m u^{\mu} \) represents the four-momentum, where
\( u^{\mu} \) is the four-velocity defined as \( u^{\mu} =
\frac{d x^{\mu}}{d \tau} \). Here, \( \tau \) is the proper time
of the test particle. Under these conditions, the Hamilton's
equations of motion can be expressed as:
\begin{equation}
\frac{d x^{\mu}}{d \zeta} \equiv m u^{\mu} = \frac{\partial H}{\partial p_{\mu}}, \qquad \frac{d p_{\mu}}{d \zeta} = -\frac{\partial H}{\partial x^{\mu}}, \label{eq:7}
\end{equation}
where the affine parameter is defined by \( \zeta = \frac{\tau}{m}\). For the present analysis, it is useful to define two new conserved quantities that arise due to the symmetries of the black
hole geometry: the specific angular momentum \( L \) and the
specific energy \( E \). These conserved quantities corresponding to
the time and azimuthal symmetries of the spacetime are defined as
\begin{equation}
\frac{p_{t}}{m} = U(r) \frac{d t}{d \tau} = -\mathcal{E}, \; \; \; \frac{p_{\phi}}{m} = r^{2} \sin^{2}\theta \, \frac{d \phi}{d \tau} =\mathcal{L},
\end{equation}
where $U(r)$ is defined in Eq. (\ref{1}) and the specific energy and specific angular momentum are given by
\( \mathcal{E} = \frac{E}{m} \) and \( \mathcal{L} = \frac{L}{m} \), respectively. The components of the four-velocity \( u^{\mu} \), namely \( u^{t} \), \( u^{\phi} \), and \( u^{r} \), satisfy the following equations of motion:
\begin{equation}
\dot{t} = \frac{\mathcal{E}}{U(r)}, \; \; \; \dot{\phi} = \frac{\mathcal{L}}{r^{2} \sin^{2}\theta}, \; \; \; \dot{r}^{2} + \left( \epsilon + \frac{\mathcal{L}^{2}}{r^{2} \sin^{2}\theta} \right) U(r) = \mathcal{E}^{2},
\end{equation}
where \( \epsilon = 1 \) corresponds to time-like particles and \(\epsilon = 0 \) corresponds to null-like particles. Here, an overdot denotes differentiation with respect to the proper time \( \tau \). In the present study, we restrict our attention to time-like particles (\( \epsilon = 1 \)). For the charged BH defined in Eq. (\ref{1}),
the Hamiltonian \cite{45} given in Eq.~(\ref{eq:6}) can be expressed as
\begin{equation}
H\; = \; \frac{1}{2} U(r) p_{r}^{2} + \frac{1}{2r^{2}} p_{\theta}^{2} +
\frac{1}{2} \frac{m^{2}}{U(r)} \left[ V_{\text{eff}}(r, \theta) -
\mathcal{E}^{2} \right],
\end{equation}
where the effective potential \( V_{\text{eff}}(r, \theta) \) is
given by
\begin{equation}
V_{\text{eff}}(r, \theta) = \left( 1 + \frac{\mathcal{L}^{2}
\csc^{2}\theta}{r^{2}} \right) U(r). \label{.}
\end{equation}

 \begin{figure}\centering
\epsfig{file=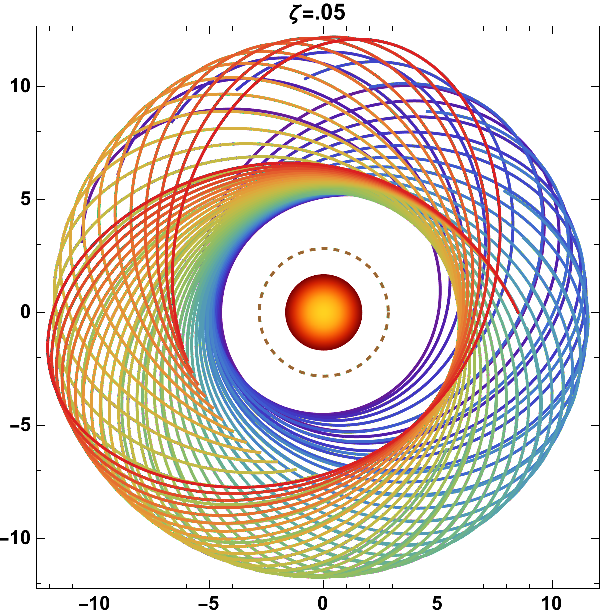,width=.4\linewidth,height=2in}\quad\epsfig{file=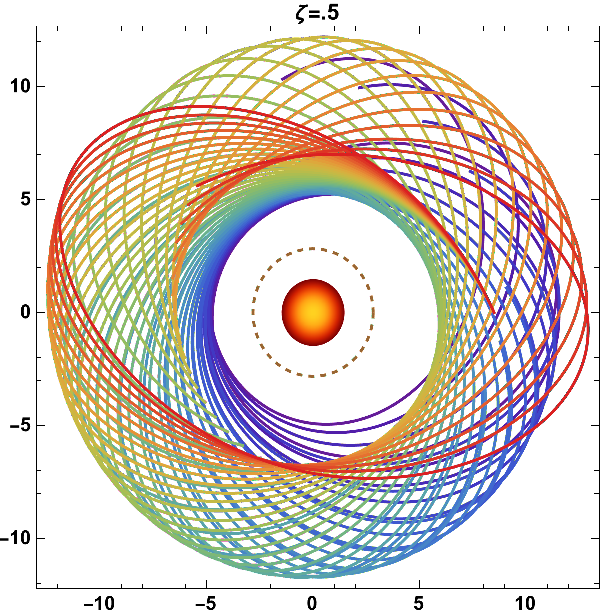,width=.4\linewidth,height=2in}\epsfig{file=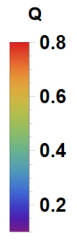,width=.06\linewidth,height=2.1in}
\epsfig{file=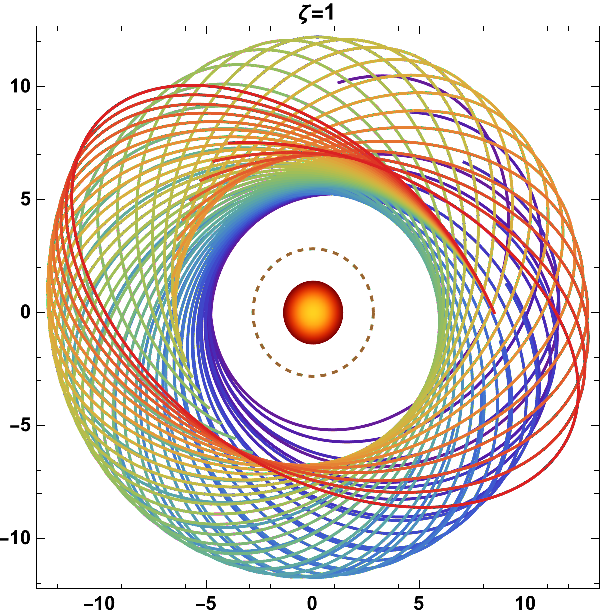,width=.4\linewidth,height=2in}\quad\epsfig{file=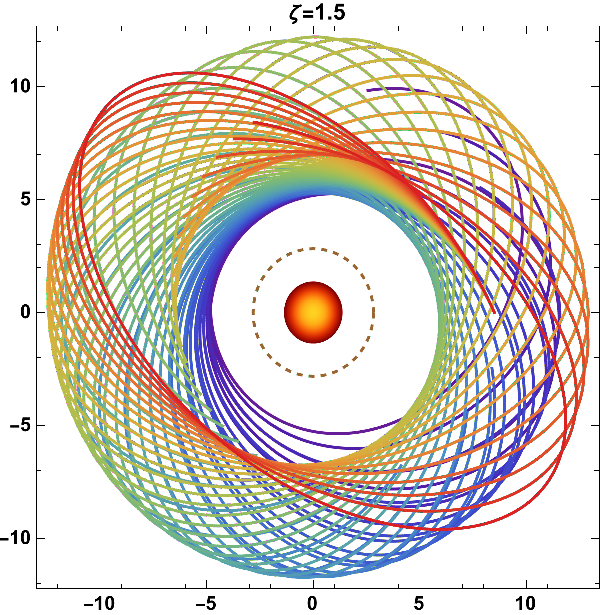,width=.4\linewidth,height=2in}\epsfig{file=al.eps,width=.06\linewidth,height=2.1in}
\caption{{\label{pt1}Particle trajectories for different values of $\zeta$.}}
\epsfig{file=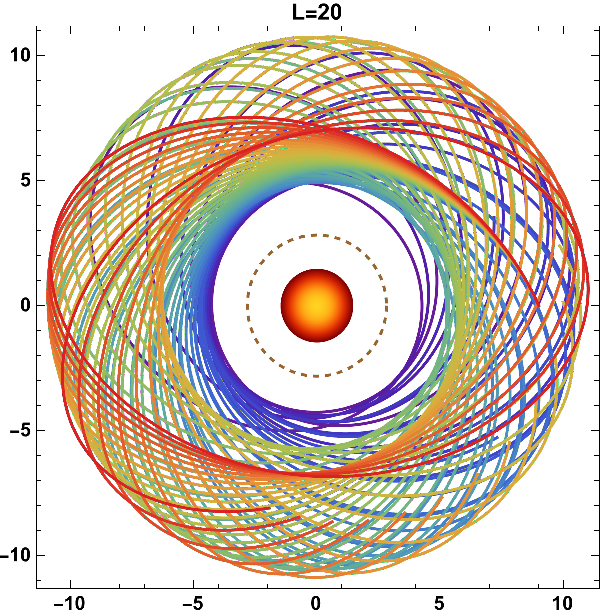,width=.4\linewidth,height=2in}\quad\epsfig{file=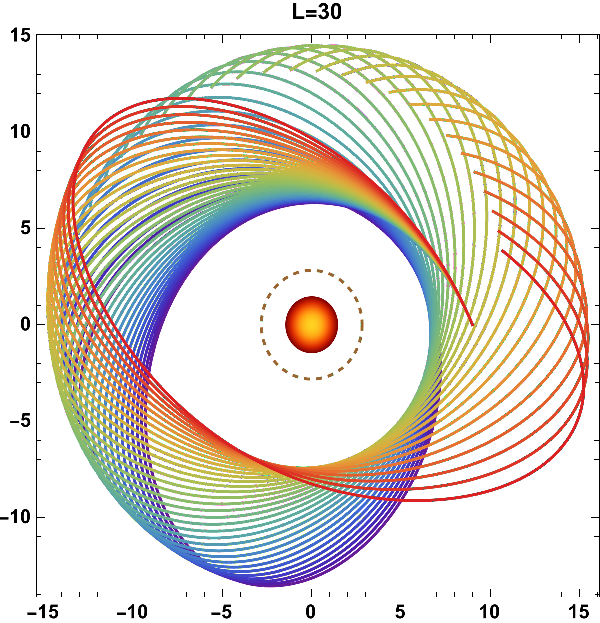,width=.4\linewidth,height=2in}\epsfig{file=al.eps,width=.06\linewidth,height=2.1in}
\caption{{\label{pt2} Particle trajectories for different values of $L$.}}
\end{figure}
For \( \zeta = Q = 0 \), the effective potential given in
Eq. (\ref{.}) reduces to the Schwarzschild case. The particle trajectories illustrated in Figs. (\ref{pt1}) and (\ref{pt2})  describe the dynamical influence of the model parameters on the orbital motion around the BH spacetime. In  Figs. (\ref{pt1}), the variation of the parameter $\zeta$ significantly modifies the effective gravitational potential governing the motion. For smaller values of $\zeta$, particles experience stronger inward attraction, leading to tightly wound precessing bound orbits near the compact object. As $\zeta$ increases, the modification in the spacetime curvature weakens the effective trapping region and the trajectories become comparatively wider, indicating an outward shift of stable orbital configurations. The color bar represents the variation of the electric charge $Q$, which directly affects the repulsive contribution of the electromagnetic field. Increasing $Q$ enhances the effective centrifugal barrier and reduces the strength of gravitational capture, producing larger orbital radii and stronger perihelion precession. Lower values of $Q$ allow particles to penetrate deeper into the gravitational well, generating tightly spiraling quasi-bound trajectories. Figure \ref{pt2} demonstrates the role of angular momentum $L$ in determining orbital stability. For relatively small angular momentum, particles undergo strong deflection and multiple precessions close to the horizon due to insufficient centrifugal support. However, increasing $L$ strengthens the centrifugal barrier, resulting in extended and more stable bound orbits at larger radial distances. These results collectively show that the interplay between the geometric parameter $\zeta$, angular momentum $L$, and charge $Q$ governs the transition between tightly bound motion, quasi-bound precessing trajectories, and wider stable orbital configurations around the BH.

\section{Effective Potential}\label{H}
The dynamics of test particle around the considered BH are examined by the effective potential obtained from the underlying spacetime geometry. The location and nature of its extrema determines possible particle trajectories and the existence and stability of circular orbits  \cite{38,39}. Moreover, its extrema helps to identify the chracteristic radii such as the photon sphere and the ISCO. For the proposed BH, the effective potential expressed as
\begin{equation}\label{Vp}
V_{\text{eff}} (r)=\left( 1 + \frac{Q^2}{{r}^2} - \frac{2M}{r} + \frac{8}{3}
P \pi {r}^2 + \frac{2 Q^2 \ln (16 Q \zeta)}{3 r \sqrt{Q \zeta}}
\right) \left( 1 + \frac{\mathcal{L}^2}{{r}^2 \sin^2 \theta} \right).
\end{equation}
The effective potential \(V_{\text{eff}}\) depends on the
the angular harmonic index \(\mathcal{L}\), effective BH charge \(Q\) and \(\zeta\).  Since the thermodynamic properties, such as the heat capacity and stability, are influenced by the effective potential, we first examine its dependence on these parameters.

\begin{figure}[h!]
    \centering
    \begin{subfigure}[b]{0.32\textwidth}
        \centering
        \includegraphics[width=\textwidth]{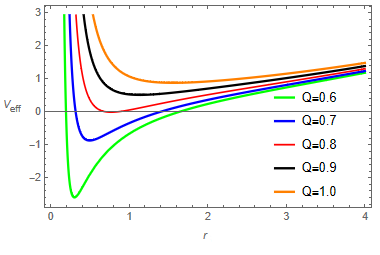}\label{veff1}
    \end{subfigure}
    \hfill
    \begin{subfigure}[b]{0.32\textwidth}
        \centering
        \includegraphics[width=\textwidth]{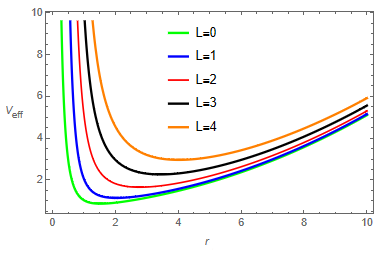}\label{veff2}
    \end{subfigure}
    \hfill
    \begin{subfigure}[b]{0.32\textwidth}
        \centering
        \includegraphics[width=\textwidth]{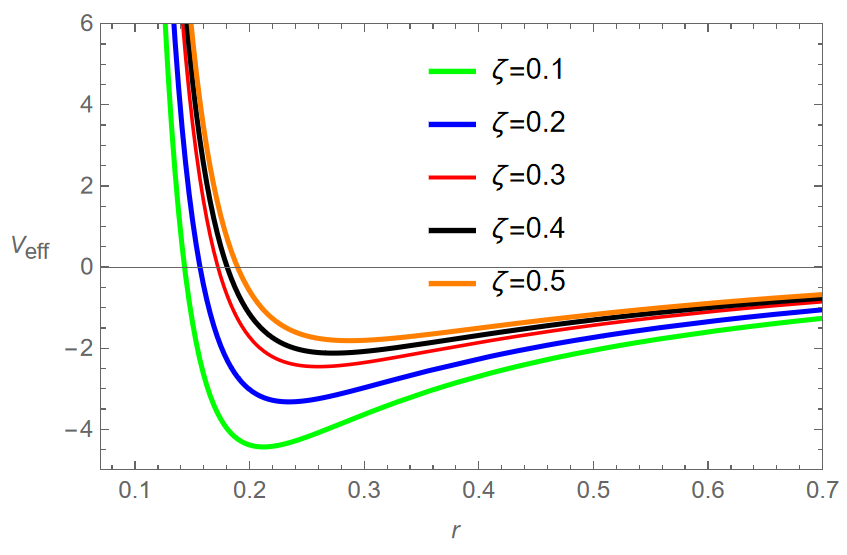}\label{veffZ}
    \end{subfigure}
    \caption{\scriptsize The plots illustrate how the effective potential \(V_{\text{eff}} \) of a BH system varies with the horizon radius \( r \) for different physical parameters: charge \( Q \), angular momentum \(\mathcal{L}\), and \( \zeta \).}\label{Veff}
\end{figure}

Fig. \ref{Veff} present the variation of \(V_{\text{eff}}\) with the radial coordinate \(r\) for selected values of \(Q\), \(\mathcal{L}\) and \(\zeta\). These plots shows the effective potential $V_{\text{eff}}$ of a test particle changes with the $r$ when varying the black-hole parameters $Q$, $\mathcal{L}$, and $\zeta$. In the first graph, for lower values of charge \(Q=0.6\) and \(Q=0.7\), the deeper potential wells indicate stronger stable bound states, suggesting that particles can remain trapped in stable orbits near the BH. In the second graph, an increase in angular momentum \(\mathcal{L}\) raises the effective potential and strengthens the centrifugal barrier near smaller \(r\), preventing particles with larger angular momentum from approaching the BH closely. For all values of \(\zeta\), the effective potential \(V_{\text{eff}}\) exhibits local minima, indicating the existence of stable circular orbits around the BH in third graph. Smaller values of \(\zeta\) corresponds to deeper potential wells, implying stronger binding of the particle and greater orbital stability. At smaller values of \(r\), the sharp increase in \(V_{\text{eff}}\) represents a strong potential barrier. Meanwhile, circular orbits are confined to the equatorial plane (i.e. at \( \theta = \pi / 2 \). These orbits are determined by the following conditions:
\begin{equation}\label{V''}
V_{\text{eff}}(r) = \mathcal{E}^2, \qquad
\frac{dV_{\text{eff}}(r)}{dr} = 0,
\end{equation}
By solving Eq. (\ref{V''}), we obtain the angular momentum equation, represents the specific angular momentum required for a particle to maintain a circular orbit around a charged BH in a spacetime influenced by pressure and quantum corrections.
\begin{equation}
\mathcal{L} \;=\; \frac{r}{\sqrt{3}} \left[{ \frac{(-3Q^{2}+3r+8P\pi
r^{4})\zeta \;-\; Q r \sqrt{Q\zeta}\,\ln(16Q\zeta)} {\displaystyle
\zeta\!\left(2Q^{2}+r(r-3)+\frac{Q^{2}r\ln(16Q\zeta)}{\sqrt{Q\zeta}}\right)}}\right]^\frac{1}{2}.
\end{equation}

\begin{figure}[h!]
    \centering
    \begin{subfigure}[b]{0.45\textwidth}
        \centering
        \includegraphics[width=\textwidth]{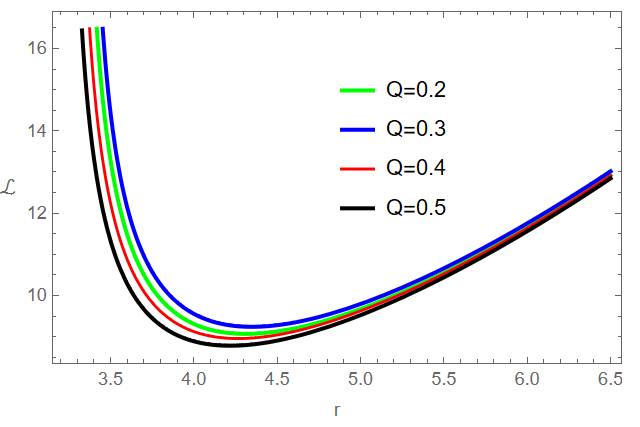}\label{L1}
    \end{subfigure}
    \begin{subfigure}[b]{0.45\textwidth}
        \centering
        \includegraphics[width=\textwidth]{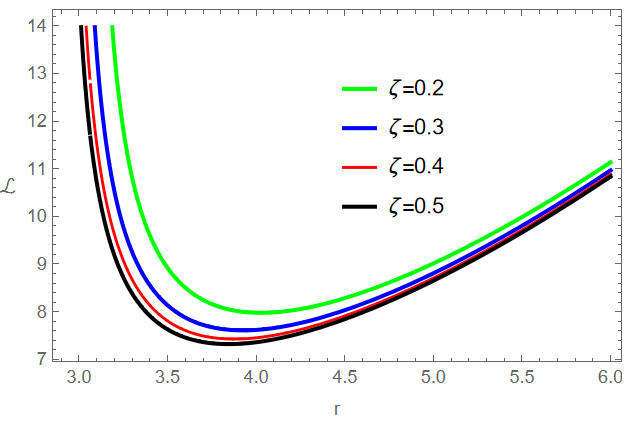}\label{L2}
    \end{subfigure}
    \caption{\scriptsize The radial variation of the angular momentum \( \mathcal{L} \)}\label{L}
\end{figure}
Fig. \ref{L} represents the graphical behavior of $\mathcal{L}$ illustrating the variations of the angular momentum $\mathcal{L}$ with the radial coordinate $r$ for different values of the charge $Q$ and the parameter $\zeta$. Both plots exhibits a characteristic U-shaped profile, indicating the existence of a stable equilibrium configuration. Similar qualitative behavior was observed when analyzing the effect of the parameter $\zeta$ on $\mathcal{L}$. In this case, $\mathcal{L}$ likewise decreases rapidly for small value of $r$, reaches a minimum, and then increases for larger values of $r$. Overall, both parameters $Q$ and $\zeta$ significantly influence the magnitude and location of the minimum of $\mathcal{L}$, while preserving the same qualitative dependence on the horizon radius, indicating the presence of a characteristic scale in the system.

\section{Innermost Stable Circular Orbits}\label{Q}

The minimum and maximum points of the effective potential correspond to the locations of stable and unstable circular orbits, respectively. In Newtonian mechanics, an effective potential always admits an innermost stable circular orbit (ISCO), representing the smallest possible radius of a stable orbit for a given angular momentum. However, in the relativistic framework, the structure of the effective potential becomes more complex due to additional factors such as the particle's angular momentum and other physical parameters. Consequently, for each value of angular momentum, the effective potential in general relativity may exhibit two extrema for particles approaching the Schwarzschild BH. To determine the ISCO, the value of the angular momentum plays a significant role. The conditions defining the ISCO are given by
\begin{equation}
V_{\text{eff}}(r) = \mathcal{E}^2, \qquad
\frac{dV_{\text{eff}}(r)}{dr} = 0, \qquad
\frac{d^2V_{\text{eff}}(r)}{dr^2} = 0,
\end{equation}
The corresponding radii of the ISCOs for a BH with are obtained by solving the above equations.
\begin{equation}
 \frac{2 \left( \mathcal{L}^2 \left( 30 Q^2 + 9(-4 + r)r \right) + r^2 \left( 9 Q^2 - 6r + 8 P \pi r^4 \right) 
+ \frac{2 Q^2 r (6 \mathcal{L}^2 + r^2) \ln (16 Q \zeta)}{\sqrt{Q \zeta}} \right)}{3 r^6}  =0 .
\end{equation}

\begin{figure}[h!]
    \centering
\includegraphics[width=8cm]{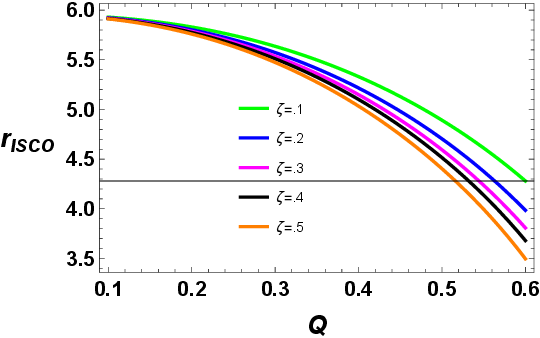}
        \includegraphics[width=8cm]{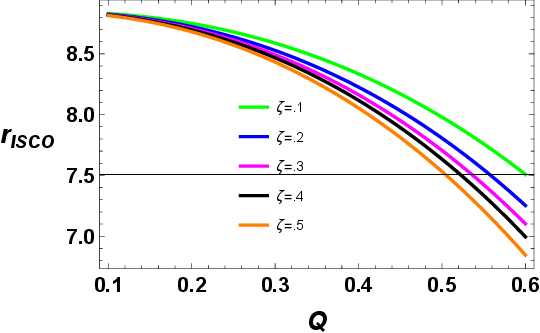}
 \caption{ Behavior of ISCO radius by varying $\zeta$ with $M=1$ (left plot) $M=1.5$ (right plot) for $L=200$. }\label{fa1}
    \centering
\includegraphics[width=8cm]{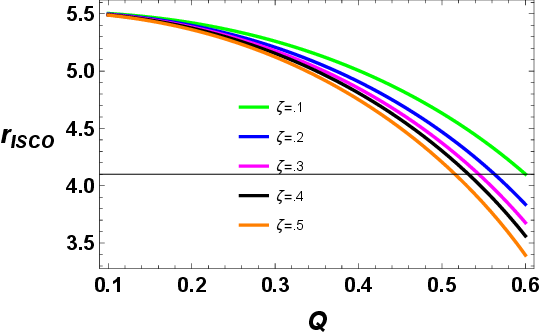}
        \includegraphics[width=8cm]{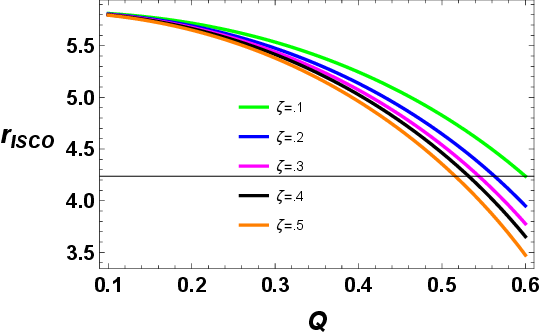}
 \caption{ Behavior of ISCO radius by varying $\zeta$ with $L=50$ (left plot) $L=100$ (right plot) for $M=1$.  }\label{fa2}
\end{figure}

We investigate the impact of BH parameters on the ISCO radius versus $Q$ as shown in Figs.
(\ref{fa1}) and (\ref{fa2}). It is found that the ISCO radius is decreases by increasing the BH charge while for massive BHs, the ISCO radius increases see Fig.
(\ref{fa1}). It is noted that the nonlinear electromagnetic parameter also reduces the ISCO radius. We also observe the impact of cosmological constant on the behavior ISCO radius as depicted in Fig. (\ref{fa2}). 

\section{Effective force}\label{I}

The motion of a particle, whether it falls towards the BH or escapes away from it, is determined by the effective force acting on it \cite{46}. This force is defined as the negative derivative of the effective potential, which incorporates both the attractive gravitational pull and the repulsive contribution arising from the BH's charge (or from angular momentum in the rotating case). In general, it is given by
\begin{equation}
    F_{\text{eff}}(r,Q) = -\frac{d}{dr} V_{\text{eff}}(r,Q).
\end{equation}
The sign and magnitude of $F_{\text{eff}}$ determine whether the particle is attracted towards the BH or repelled away from it. Points where $F_{\text{eff}}=0$ correspond to possible equilibrium radii, the stability of which depends on the local behavior (slope) of $F_{\text{eff}}(r)$ near these points \cite{51}. Using Eq. (\ref{Vp}), the resulting effective force is:
\begin{equation}
\begin{split}
    F(r) =-\left( \frac{-3 \mathcal{L}^2 \left( 2 Q^2 + (-3 + r)r \right)+ {r}^2 \left( -3 Q^2 + 3r + 8 P \pi {r}^4 \right)- \frac{Q^2 r(3 \mathcal{L}^2 + {r}^2) \ln (16 Q \zeta)}{\sqrt{Q \zeta}}}{3 {r}^5}\right).
\end{split}
\end{equation}
\begin{figure}[h!]
    \centering
    \begin{subfigure}[b]{0.32\textwidth}
        \centering
        \includegraphics[width=\textwidth]{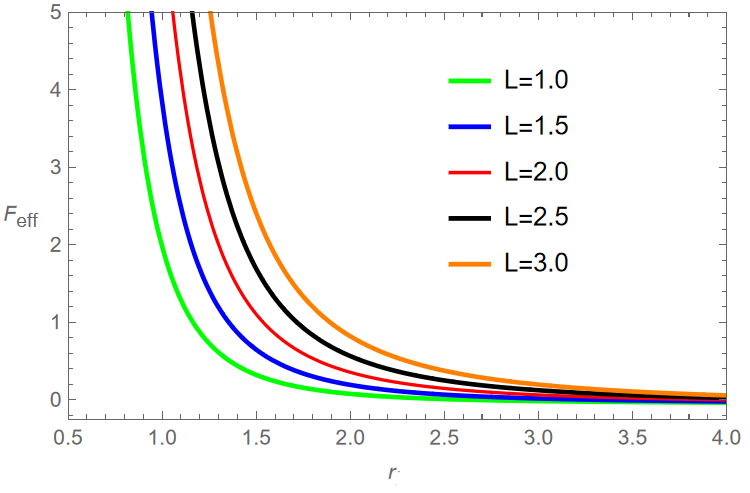}\label{F1}
    \end{subfigure}
    \hfill
    \begin{subfigure}[b]{0.32\textwidth}
        \centering
        \includegraphics[width=\textwidth]{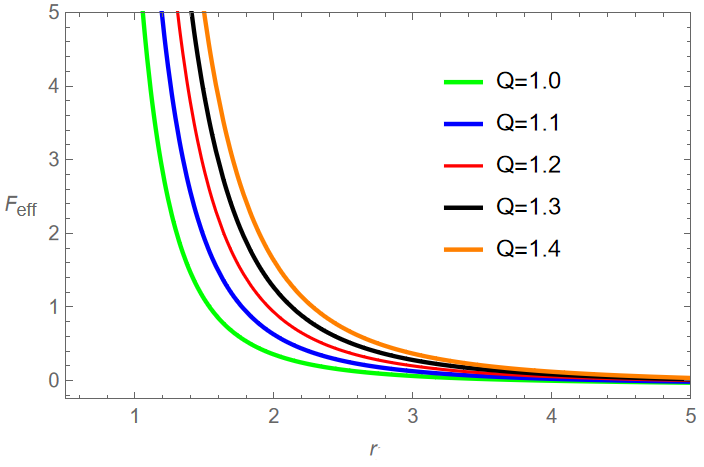}\label{F2}
    \end{subfigure}
    \hfill
    \begin{subfigure}[b]{0.32\textwidth}
        \centering
        \includegraphics[width=\textwidth]{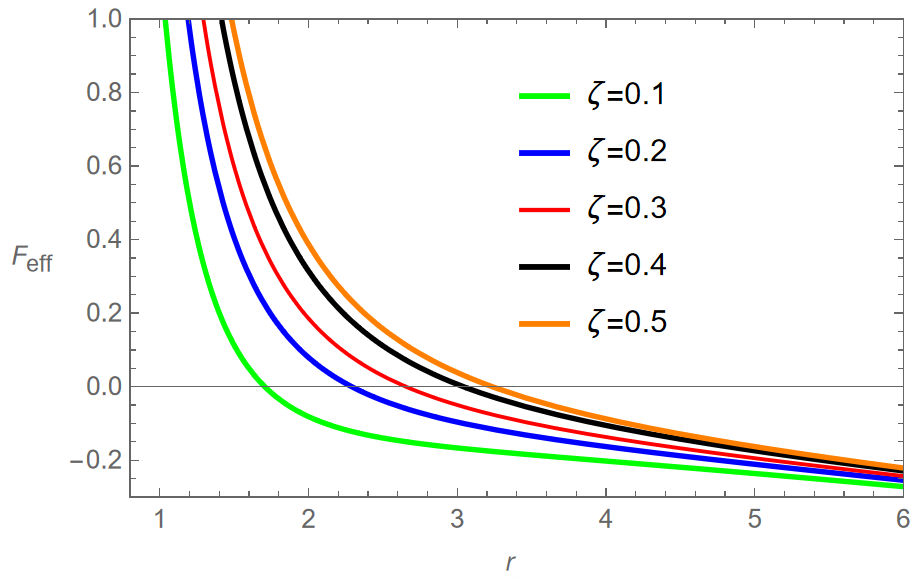}\label{F3}
    \end{subfigure}
    \caption{\scriptsize  Plots illustrate how the effective force \(F_{\text{eff}} \) of a BH system varies with the horizon radius \( r \) involving different physical parameters: charge \( Q \), angular momentum \(\mathcal{L}\), and parameter \( \zeta \).}\label{F}
\end{figure}

Figure \ref{F} illustrates the graphical characteristics of effective force $F_{\text{eff}}$ associated with motion of test particle as a function of the radial coordinate $r$, considering various values for the angular momentum $\mathcal{L}$, electric charge $Q$, and parameter $\zeta$, which shows the effective force is strongest near the event horizon and gradually weakens as $r$ increases, and eventually approaches zero at large radial distances. The position of the zero crossing and the overall magnitude of $F_{\text{eff}}$ are strongly influenced by $\zeta$. Increasing $\zeta$ shifts the curves upward, causing the zero-crossing point to occur at larger values of $r$. Accordingly, for larger $\zeta$, the repulsive regime persists over a wider range of horizon radii. The graph demonstrates that the effective force acting on a test particle remains constant. However, the rotational and electromagnetic effects may strengthens the effective force.

\section{Harmonic oscillations as perturbation of circular orbits}\label{HM}
The epicyclic frequencies of neutral test particle are determined by analyzing small perturbations around stable circular orbit. These frequencies are calculated by local observer are stated as
\begin{equation}
\omega_{r}^{2} = -\frac{1}{2} \frac{\partial^{2} V_{\text{eff}}(r, \theta)}{\partial r^{2}}, \; \; \; \omega_{\theta}^{2} = \frac{1}{2} g_{rr}(r, \theta) \frac{1}{r^{2}} \frac{\partial^{2} V_{\text{eff}}(r, \theta)}{\partial \theta^{2}}, \; \; \; \omega_{\phi} = \frac{d\phi}{d\tau}.
\end{equation}
The radial $\left(\omega_{r}\right)$, latitudinal $\left(\omega_{\theta}\right)$, and azimuthal/orbital test particle frequencies $\left(\omega_{\phi}\right)$ are given by the general form:

\begin{equation}\label{J1}
\omega_{r}^{2} ={\textstyle{-\frac{ \Big( J \zeta + Q r (9 Q^{2} + r (-12 + r - 40 P \pi r^{3})) \sqrt{Q\zeta} \log(16 Q \zeta) + 2 Q^{3} r^{2} \ln^{2}(16 Q \zeta) \Big)}{ 3 r^{4} \left( 2 Q^{2} \zeta + (-3 + r) r \zeta + Q r \sqrt{Q\zeta} \ln (16 Q \zeta) \right) }}},
\end{equation}
where $J$ expressed in Appendix in Eq. (\ref{j}).
\begin{equation}\label{Omg1}
\omega_{\theta}^{2} =\frac{ \left( (-3 Q^{2} + 3 r + 8 P \pi r^{4})
\zeta - Q r \sqrt{Q \zeta} \ln (16 Q \zeta) \right) }{ 3 r^{2} \zeta
\left( 2 Q^{2} + (-3 + r) r + \frac{Q^{2} r \ln(16 Q
\zeta)}{\sqrt{Q \zeta}} \right) },
\end{equation}

\begin{equation}\label{Omg2}
\omega_{\phi}^{2} =\frac{ \left( (-3 Q^{2} + 3 r + 8 P \pi r^{4})
\zeta - Q r \sqrt{Q \zeta} \ln (16 Q \zeta) \right) }{ 3 r^{2} \zeta
\left( 2 Q^{2} + (-3 + r) r + \frac{Q^{2} r \ln (16 Q
\zeta)}{\sqrt{Q \zeta}} \right) }.
\end{equation}
 It is clear from Eqs. (\ref{Omg1}) and (\ref{Omg2}) that $\omega_{\theta}=\omega_{\phi}$.

\subsection{Frequencies measured by distant observer}
The angular frequencies as measured by a static distant observer
$(\Omega)$ are given by
\begin{equation}
\Omega_\alpha = {\omega_\alpha} \frac{d\tau}{dt}.
\end{equation}
The quantity ${d\tau}/{dt}$ represents the redshift coefficient. When the frequencies of small harmonic oscillations measured by a distant observer are expressed in physical units, the corresponding dimensionless frequencies must be multiplied by the factor ${c^{3}}/{G M}$. Hence, the frequencies of neutral particle oscillations as detected by a distant observer are given by:
\begin{equation}
\nu_{j} = \frac{1}{2\pi}\,\frac{c^{3}}{G M}\,\Omega_{j}\;[\text{Hz}].
\end{equation}
Here, $j \in \{r, \theta, \phi\}$, and 
$\Omega_{r}$, $\Omega_{\theta}$, and $\Omega_{\phi}$ denote 
the dimensionless radial, latitudinal, and azimuthal 
angular frequencies measured by a distant observer, given by
\begin{equation}
\Omega_r^2 = \textstyle{\frac{-J\zeta- Q r \left(9 Q^{2} + r(-12 + r - 40 P \pi r^{3})\right)\sqrt{Q \zeta}\,\ln(16 Q \zeta)- 2 Q^{3} r^{2}\ln^{2}(16 Q \zeta)}{3 r^{2} (\mathcal{L}^{2} + r^{2})\,\left(2 Q^{2}\zeta + (-3 + r) r \zeta+ Q r \sqrt{Q \zeta}\,\ln (16 Q \zeta)\right)}},
\end{equation}

\begin{equation}
\Omega_\theta^2=\frac{\left(-3 Q^{2} + 3 r + 8 P \pi r^{4}\right)\sqrt{Q \zeta}- Q^{2} r \ln (16 Q \zeta)}{3 (\mathcal{L}^{2} + r^{2})\left[\left(2 Q^{2} + (-3 + r) r\right)\sqrt{Q \zeta}+ Q^{2} r \ln (16 Q \zeta)\right]},
\end{equation}

\begin{equation}
\Omega_\phi^2=\frac{\left(-3 Q^{2} + 3 r + 8 P \pi r^{4}\right)\sqrt{Q \zeta}- Q^{2} r \ln (16 Q \zeta)}{3 (\mathcal{L}^{2} + r^{2})\left[\left(2 Q^{2} + (-3 + r) r\right)\sqrt{Q \zeta}+ Q^{2} r \ln (16 Q \zeta)\right]}.
\end{equation}

\subsection{Periastron Precession}
Now, we investigate the periastron precession frequency $\Omega_{p}$ of a neutral test particle, non-rotating, scalar charged BH in which the particle experiences a small perturbation at an angle of $\pi/2$ relating to the equatorial plane. A slight displacement from its stable circular orbit produces oscillatory motion about the equilibrium position, characterized by the radial epicyclic frequency $\Omega_{r}$. Since $\Omega_{r}$ differs from the $\Omega_{\phi}$, the $\Omega_{P}$ is obtained from the relation $\Omega_{P} = \Omega_{\phi} - \Omega_{r}$.
\begin{figure}[h!]
    \centering
    \begin{subfigure}[b]{0.32\textwidth}
        \centering
        \includegraphics[width=\textwidth]{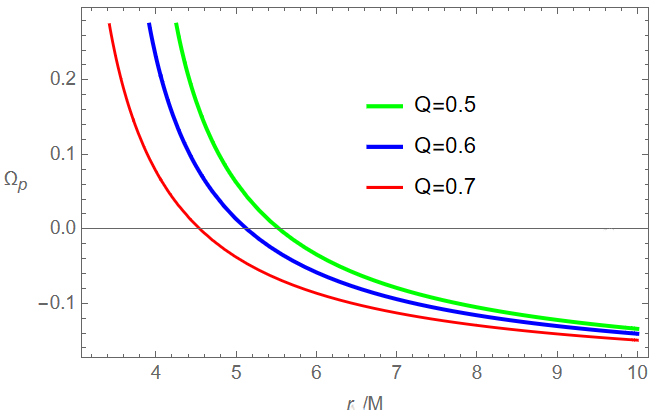}\label{S1}
    \end{subfigure}
    \hfill
    \begin{subfigure}[b]{0.32\textwidth}
        \centering
        \includegraphics[width=\textwidth]{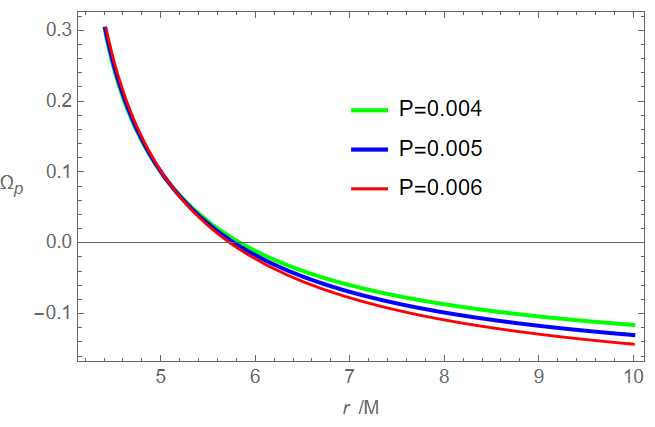}\label{S2}
    \end{subfigure}
    \hfill
    \begin{subfigure}[b]{0.32\textwidth}
        \centering
        \includegraphics[width=\textwidth]{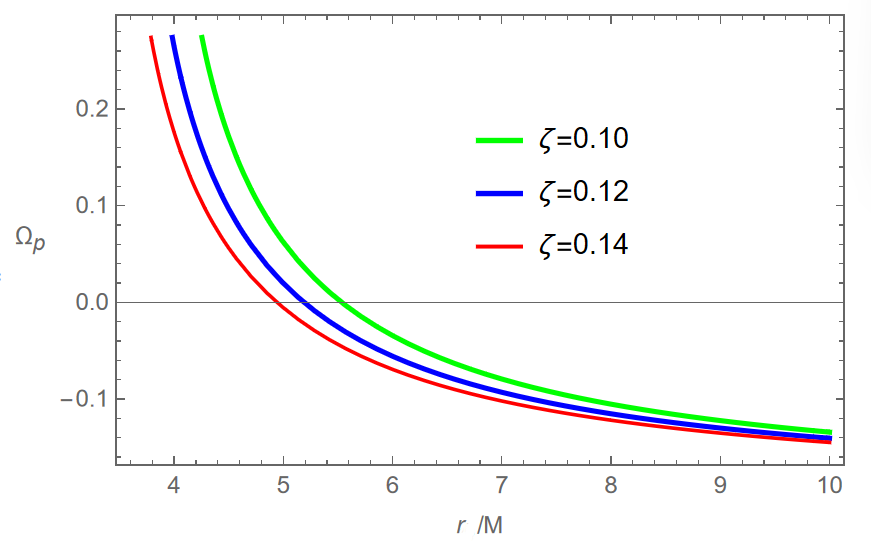}\label{S3}
    \end{subfigure}
    \caption{\scriptsize Radial profiles of the $\Omega_{p}$ for a charged BH with different $Q$ and the thermodynamic pressure $P$.}\label{S}
\end{figure}
The graphical behavior of $\Omega_{P}$ is shown in Fig. \ref{S}, which illustrate the influence of BH parameters on $\Omega_{p}$ for particles orbiting a charged BH. The first graph shows that increasing the charge $Q$ shifts the curves downward, indicating that stronger electromagnetic effects reduces the precession rate and alters the radius at which $\Omega_{p}$ changes sign. The second graph illustrates the thermodynamic pressure $P$, where higher pressure values slightly suppress $\Omega_{p}$ across the entire radial range. The third graph illustrates the variation of the quantity $\Omega_p$ for different values of the parameter $\zeta$. For all cases, $\Omega_p$ decreases monotonically as $r$ increases, it is positive at small radii, crosses zero at an intermediate radius, and becomes negative at larger radii. Increasing $\zeta$ shifts the curve downward, such that for any fixed value of $r$, a larger $\zeta$ corresponds to a smaller value of $\Omega_p$. Consequently, the zero-crossing point of $\Omega_p$ occurs at smaller $r$ as $\zeta$ increases.

\section{Center of mass energy}\label{N}
To examine the substantial energy obtained during the collision of two neutral test particles in a strong gravitational field (especially near a BH), the Centre-of-Mass (CM) energy is applied which gives the fundamental knowledge about the dynamics of the collision and determines how the BH’s physical parameters, such as mass, charge, and angular momentum, affect the energy of colliding particles. The particles considered with the same rest mass but different four-velocities, traveling in the same plane, the CM energy is defined as
\begin{equation}
\frac{\sqrt{E_{\mathrm{cm}}}}{2 m_{0}}
    = \sqrt{1 - g_{\mu\nu} u^{\mu} u^{\nu}} ,
\end{equation}
where the four-velocity components of the $i$th particle are
\begin{equation}
u^{\mu}_{(i)} =\left(\frac{\mathcal{E}_{i}}{U(r)},\; -Y_{i}, \; 0,\; \frac{\mathcal{L}_{i}}{r^{2}}\right),
\quad\;and\;\quad
Y_{i} =\sqrt{\mathcal{E}_{i}^{2}-V_eff
},
\qquad i = 1,2.
\end{equation}

with $\mathcal{E}_{i}$ represents the conserved energy and $\mathcal{L}_{i}$ represents the angular momentum of each particle. For two particles moving in the gravitational field of a charged BH, the CM energy can be expressed  as

\begin{equation}
\frac{\sqrt{E_{\mathrm{cm}}}}{2 m_{0}}=\sqrt{1+ \frac{\mathcal{E}_{1}\mathcal{E}_{2}}{U(r)}- \frac{Y_{1} Y_{2}}{U(r)}- \frac{\mathcal{L}_{1} \mathcal{L}_{2}}{r^{2}}},
\end{equation}
Here, $U(r)$ is using from Eq. (\ref{1}) as
\begin{equation}\label{Y}
\frac{\sqrt{E_{\mathrm{cm}}}}{2 m_{0}}=\left[{1-\frac{\mathcal{L}_{1} \mathcal{L}_{2}}{r^{2}}+\frac{\mathcal{E}_{1}\mathcal{E}_{2}}{ \left( 1 + \frac{Q^{2}}{r^{2}} - \frac{2M}{r} + \frac{8}{3} P\pi r^{2} + \frac{2 Q^{2} \ln\!\left(16 Q \zeta\right)}{3 {r} \sqrt{Q\zeta}} \right)}- \frac{Y_{1} Y_{2}}{ \left( 1 + \frac{Q^{2}}{r^{2}} - \frac{2M}{r} + \frac{8}{3} P\pi r^{2} + \frac{2 Q^{2} \ln\!\left(16 Q \zeta\right)}{3 {r} \sqrt{Q\zeta}} \right)}}\right]^\frac{1}{2}.
\end{equation}
Here, $Y_{1}, Y_{2}$ are defined in Appendix.

\section{Conclusion}\label{J}

In this work, we have presented a comprehensive investigation of the thermodynamic properties and orbital dynamics of logarithmically corrected charged anti de-Sitter black hole within the framework of Sharma-Mittal generalized entropy which is characterized by the three parameters \(\alpha\), \(\beta\), and \(\gamma\). We have explored the modifications induced by this entropy on the thermodynamic properties and phase structure of the black hole configuration. By employing Duan’s $\Phi$-mapping formalism, we have investigated the topological structure of the thermodynamic critical points in both canonical and mixed ensembles, thereby establishing a complementary perspective on black hole thermodynamic stability. We have determined different topological properties in the canonical and mixed ensembles by building the corresponding vector field, finding its zeros and computing its winding numbers. We derive explicit expressions for the mass $M$, temperature $T$, Gibbs free energy $F$, heat capacity $C$, effective potential $V_{\text{eff}}(r)$, and effective force while treating the negative cosmological constant as thermodynamic pressure. The results demonstrate that the entropy parameters exert a strong influence on BH thermodynamics, significantly modifying the behavior of the mass, temperature, and Gibbs free energy, which is in accordance with the previous findings in \cite{53,OME}. The thermodynamic topological framework results provide (W=-1), while the thermodynamic defect provided as (W=+1) for the non vanishing contribution. 

We further investigate critical points and phase structures revealing their corresponding topological classes and establishing how the winding number and topological charge act as robust global invariants to characterize BH stability and distinguish standard critical phenomena from entropy-corrected novel phase transitions (see Figs. \ref{fig.N}, \ref{O1}, \ref{O2}, \ref{O3}). The analytical criticality conditions further establish the critical horizon radius, pressure, and temperature, as obtained in Eq. (\ref{cp}). Heat capacity analysis further supports these classifications, where its sign determines local thermodynamic local stability and its divergences, consistent with Eq. (\ref{12}), signal phase transitions that closely align with variations in the effective potential and the behavior of isobaric curves near the critical point, while the effective potential V(r) offers a complementary description by illustrating how BH parameters shape the horizon geometry and govern the surrounding thermodynamic phase structure. We conducted an analysis of the heat capacity as a function of the horizon radius and identified regions of thermodynamic stability \( (C > 0) \) and instability \( (C < 0) \). In Fig. \ref{hc}, Positive heat capacity confirms the thermal stability of black hole. Our results show that the electric charge, angular momentum, and logarithmic coupling parameter \(\zeta\) have a significant effect on the orbital dynamics. In particular, the smaller the value of \(\zeta\), the tighter and more strongly precessing will be the trajectories, while the higher the value, the wider will be the orbits. Likewise, a change in the black hole charge alters the effective potential and makes it more repulsive, which leads to orbits at a larger radius. In Fig. \ref{Veff}, the logarithmic term and NED corrections significantly change the effective potential. In Fig. \ref{fa1}, \ref{fa2}, we discuss the location of the ISCO orbital oscillation frequencies. Finally, we have derived an analytical expression for the center of mass energy of two colliding neutral test particles, providing further insight into particle collision processes. \\

The geodesic analysis reveals how the spacetime parameters govern orbital stability, the location of the ISCO, epicyclic oscillations, and particle collision energetics. Our study offers a more comprehensive understanding of the interplay between black hole thermodynamics and strong-field particle dynamics. Additionally, we provide a comparison of our current study with the literature in recent years  \cite{15}, \cite{16}, \cite{38}, \cite{OME}, \cite{46}. For Future work, we may enhance the observational implications of these findings through black hole shadows, quasinormal modes, and other astrophysical signatures, as well as extension of analysis to rotating, higher-dimensional, and quantum-corrected black hole solutions.

\section{Appendix}
For the clarity and to avoid repetition in the main work, some of the abbreviated terms used herein are summarized below. The parameter $A$ appearing in Eq. (\ref{phi1}) is defined as follow
\begin{equation}\label{A}
A=(6 Q^2 - r_h^2) \beta^2 + 2 \pi r_h^2 \alpha \beta \left( Q^2 (14 -
8\beta) + r_h^2 (-3 + 2\beta) \right) - \pi^2 r_h^4 \alpha^2 (-3 +
2\beta) \left( r_h^2 (1 - 2\beta) + Q^2 (2 + 4\beta) \right).
\end{equation}
The parameter $B$ and $C$ used in Eq. (\ref{phi2}) is defined as
\begin{equation}\label{b}
B=\beta \left( -2 r_h \phi_e^2 + 3 \sqrt{r_h \zeta \phi_e} (1 +
\phi_e^2) \right)+ \pi r_h^2 \left( 2 r_h \alpha (-3 + 2\beta) \phi_e^2
+ 3 \sqrt{r_h \zeta \phi_e} \left( -8 P \beta + \alpha (-1 + 2\beta)(1
+ \phi_e^2) \right) \right).
\end{equation}

\begin{equation}\label{c}
C=\beta + \pi r_h^2 \alpha (-3 + 4\beta).
\end{equation}
In Eq. (\ref{N1}), the parameter $D$ is introduced and defined as
\begin{equation}\label{d}
D=2\phi_e \sqrt{r_h\zeta \phi_e} + 3\zeta (1+8P\pi r_h^2 + \phi_e^2) +
3\phi_e \sqrt{r_h\zeta \phi_e}\ln(16 r_h \zeta \phi_e),
\end{equation}
The parameter $E$ appearing in Eq. (\ref{E1}) is defined as 
\begin{equation}\label{e}
E=- \frac{12 \pi r_h \alpha \left(1 + \frac{\pi r_h^2 \alpha}{\beta}
\right)^{-1 + \beta}}{\gamma \tau} - 3 \phi_e^2 + \frac{2 \phi_e
\sqrt{r_h \zeta \phi_e}}{\zeta}.
\end{equation}
The parameter $J$, which enters in Eq. (\ref{J1}), is defined as follows
\begin{equation}\label{j}
J= (12 Q^{4} - 3 Q^{2} r (9 +
32 P \pi r^{3}) + r^{2} (18 - 3 r + 120 P \pi r^{3} - 32 P \pi
r^{4}))
\end{equation}
Finally, the parameters $Y_1$, $Y_2$ used in Eq. (\ref{Y}), are defined as follows
\begin{equation}\label{y}
Y_1 = \left[{\epsilon_1^{2} - \left(1 + \frac{\mathcal{L}_1^{2}}{r^{2}}\right)\left(1 + \frac{Q^{2}}{r^{2}} - \frac{2M}{r} + \frac{\alpha}{r^{2}}\right)}\right]^\frac{1}{2} ,
\end{equation}

\begin{equation}\label{y0}
Y_2 = \left[{\epsilon_2^{2} - \left(1 + \frac{\mathcal{L}_2^{2}}{r^{2}}\right)\left(1 + \frac{Q^{2}}{r^{2}} - \frac{2M}{r} + \frac{\alpha}{r^{2}}\right)}\right]^\frac{1}{2} .
\end{equation}

\section*{Acknowledgements}
The work was sponsored by the National Natural Science Foundation of China (Grant Numbers: 12235007, 12375003).

\bibliographystyle{apsrev4-1}

\end{document}